\documentclass[%
 reprint,
 amsmath,amssymb,
 aps,
]{revtex4-1}

\usepackage{graphicx}% Include figure files
\usepackage{dcolumn}% Align table columns on decimal point
\usepackage{bm}% bold math
\usepackage[usenames,dvipsnames]{color}
\usepackage[normalem]{ulem}
\renewcommand{\vec}[1]{\mathbf{#1}}

\newcommand{\lo}[1]{\color{black} #1 \color{black}}

\begin{document}

\preprint{APS/123-QED}

\title{Resting and Traveling Localized States in an Active Phase-Field-Crystal Model}

\author{Lukas Ophaus}

\author{Svetlana V. Gurevich}

\author{Uwe Thiele}
 \email{u.thiele@uni-muenster.de}
 \affiliation{Institut f\"ur Theoretische Physik, Westf\"alische
Wilhelms-Universit\"at M\"unster\\ Wilhelm-Klemm-Strasse 9, 48149
M\"unster, Germany}
 \affiliation{Center of Nonlinear Science (CeNoS), Westf\"alische
Wilhelms-Universit\"at M\"unster\\ Corrensstrasse 2, 48149 M\"unster,  Germany}

\date{\today}

\begin{abstract}
The conserved Swift-Hohenberg equation (or Phase-Field-Crystal [PFC] model) provides a simple microscopic description of the thermodynamic transition between fluid and crystalline states. Combining it with elements of the Toner-Tu theory for self-propelled particles Menzel and L\"owen [Phys.\ Rev.\ Lett.\ \textbf{110}, 055702 (2013)] obtained a model for crystallization (swarm formation) in active systems. Here, we study the occurrence of resting and traveling localized states, i.e., crystalline clusters, within the resulting active PFC model. Based on linear stability analyses and numerical continuation of the fully nonlinear states, we present a detailed analysis of the bifurcation structure of periodic and localized, resting and traveling states in a one-dimensional active PFC model. This allows us, for instance, to explore how the slanted homoclinic snaking of steady localized states found for the passive PFC model is amended by activity. A particular focus lies on the onset of motion, where we show that it occurs either through a drift-pitchfork or a drift-transcritical bifurcation. A corresponding general analytical criterion is derived.

\end{abstract}

\pacs{Valid PACS appear here}% PACS, the Physics and Astronomy
\maketitle

\section{\label{sec:level1}Introduction}

Active particles like bacteria, animals or artificial micro-swimmers \cite{wadaPRL99,rappelPRL83,vicsekPRE74,sumpter} are able to transform different forms of energy into self-propelled directed motion \cite{Marchetti.RevModPhys.85,BDLR2016rmp}. They use various energy sources to drive some internal motor mechanism and re\-present out of equilibrium systems driven by a continuous energy flow. Artificial micro-swimmers, for instance, turn chemical energy \cite{howsePRL99} or radiation like light \cite{PalacciScience,jiangPRL105} or ultrasound \cite{WangACSNano2012} into an actively driven, self-propelled motion. 

Non-equilibrium systems that are composed of a large number of active particles can show fascinating collective phenomena. In particular, short- and long-range interactions between individual particles result in alignment mechanisms that can cause directional ordering (so-called polar ordering) and synchronization of the motion of self-propelled particles \cite{uchidaPRL106,golestanianSoftMatter7}. The resulting collective modes of motion are often referred to as swarming \cite{Marchetti.RevModPhys.85}. Also, vibrated granular media in confined geometries are employed as good model systems for certain aspects of collective behavior of active particles \cite{GranularExcitonsNature1996,ArTs2003pre,WHDL2013prl,NaMR2006jsme}.

\lo{Depending on the particular interactions between particles, their density and strength of driving (activity) one observes different regimes of clustering, ordering and motion that one may, in analogy to equilibrium behavior call gas, liquid, liquid-crystalline and crystalline states \cite{BDLR2016rmp,MaVC2018arpc}. Much recent attention focused on an actively driven condensation phenomenon, the motility-induced phase separation between a gaseous and a liquid state that is purely due to self-propulsion \cite{Ginot2015prx,SSWK2015prl,CaTa2015arcmp}. However, for certain particle interactions and/or at quite high densities active particles can also form crystalline ordered states, in particular, resting \cite{Thar2002,Thar2005} or traveling \cite{PalacciScience,theurkauff2012prl,LibchaberPRL2015,ginot2018aggregation} patches with nearly crystalline order \cite{ToTR2005ap}.  These ``active crystals'' \cite{MenzelLoewen,MenzelOhtaLoewenPhysRevE.89} (called ``flying crystals'' in \cite{ToTR2005ap} and ``living crystals'' in \cite{PalacciScience,MSAC2013prl,BDLR2016rmp}) have properties that differ from passive crystalline clusters \cite{speckPRL112,speckPRL110}. The activity due to self-propulsion can change the critical temperature and density at which crystallization sets in. Besides, it can induce organized translational and rotational motion \cite{theurkauff2012prl,ReBH2013pre,Ginot2015prx,ginot2018aggregation}. 
Many particle-based models are studied that show resting, traveling and rotating, active, crystalline and amorphous clusters \cite{EbEr2003pspi,REES2008epjt,MSAC2013prl,NKEG2014prl} as well as cluster-crystals \cite{Menz2013jpm,DeLH2017njp}.  For instance, a systematic study of the interplay of a short-range attraction and self-propulsion in Brownian dynamics simulations shows that clusters form at low activity (due to attraction) as well as at high activity (motility-induced) with a homogeneous active fluid phase in between \cite{ReBH2013pre}.}

\lo{There exist many continuum models for active matter \cite{ToTR2005ap,Marchetti.RevModPhys.85,Menz2015prspl,RKBH2018pre}, an important example is the Toner-Tu model of swarming \cite{ToTu1995prl,TonerTu}. It represents a generalization of the compressible Navier-Stokes equations of hydrodynamics to systems without Galilei invariance, i.e., with preferred velocities.
Recently, a simple active Phase-Field-Crystal model (aPFC) has been proposed that describes transitions between the liquid state and resting and traveling crystalline states \cite{MenzelLoewen}.
It combines elements of the Toner-Tu theory and the (passive) Phase-Field-Crystal model (PFC), an intensively studied microscopic continuum model for the dynamics of crystallization processes on diffusive time scales \cite{EmmerichPFC}.}

The Phase-Field-Crystal model was introduced by Elder and coworkers \cite{ElderGrantPRL88} and is applied for passive colloidal particles but also used for atomic systems \cite{TGTP2009prl,ERKS2012prl}. Mathematically, it corresponds to the conserved Swift-Hohenberg equation (cSH) \cite{TARG2013pre}, i.e., the counterpart with conserved dynamics (i.e., of the form of a continuity equation) of the Swift-Hohenberg (SH) equation that represents non-conserved dynamics \cite{EGUW2018springer}. The latter is the standard equation for pattern formation close to the onset of a monotonous short-wave instability in systems without a conservation law, e.g., a Turing instability in reaction-diffusion systems or the onset of convection in a B\'enard system \cite{CrossHohenberg}.  The cSH equation was first derived as the equation governing the evolution of binary fluid convection between thermally insulating boundaries \cite{KnoblochPRA1989};  in the PFC context recent derivations from classical Dynamical Density Functional Theory (DDFT) of colloidal crystallization can be found in Refs.~\cite{EmmerichPFC,ARTK2012pre}. In the course of the derivation, the one-particle density of DDFT is shifted and scaled to obtain the order parameter field of PFC. For brevity, in the following we refer to it as ``density''. Note that both, SH and PFC models, represent gradient dynamics on the same class of energy functionals \cite{EGUW2018springer}. However, in the active PFC model the coupling between density and polarization (quantified by the coupling or activity parameter) breaks the gradient dynamics structure, therefore allowing for sustained motion. Note that non-variational amendments of the standard non-conserved SH equation are also studied and can also show traveling states, though with different onset behavior \cite{KoTl2007c,HoKn2011pre,BuDa2012sjads}.

Up to now the active Phase-Field-Crystal model has mainly been employed to study the linear stability of the liquid state with respect to the development of resting and traveling crystalline patterns and in the study of domain-filling resting and traveling crystals by direct time simulations \cite{MenzelLoewen,MenzelOhtaLoewenPhysRevE.89,ChGT2016el,PVWL2018pre}. 

The main purpose of the present work is to investigate resting and traveling, periodic and localized states and the related transitions as described by the active Phase-Field-Crystal model.  
\lo{Our aim is to present a detailed analysis of the underlying bifurcation structure that can serve as reference for future similar analyses of other models describing active crystals. This shall allow one to develop a clearer understanding of observed multistabilities of states, hysteresis effects and critical threshold states for the occurrence of qualitative changes.
Here, a particular focus is on the transitions from resting to traveling states that} will turn out to occur at drift-pitchfork and drift-transcritical bifurcations. Drift-pitchfork bifurcations are widely studied in the literature and occur in many systems \cite{FaDT1991jpi,GGGC1991pra}. This includes the onset of motion of self-aggregating membrane channels \cite{LeNH2006prl}, drifting liquid column arrays \cite{BrFL2001el}, chemically-driven running droplets \cite{JoBT2005epje} and traveling localized states in reaction-diffusion systems~\cite{SOBP_PRL97,PiPRL01,driftbif_gurevich}. The onset of motion for localized structures is studied, for instance, in Refs.~\cite{krischerPRL73,Or-Gui.PhysRevE.57,baerPRE64,akhmedievPRE53} while Refs.~\cite{MenzelOhtaLoewenPhysRevE.89,OSIPOV1996,PVWL2018pre} focus on domain-filling patterns. 

In the PFC and aPFC models, spatially localized states correspond to finite crystalline patches (i.e., patches of periodic states) that coexist with a liquid background (i.e., a homogeneous state). A great variety of resting localized states has been analyzed in detail for the PFC model in Ref.~\cite{TARG2013pre} where detailed bifurcation diagrams are given in the case of one spatial dimension \lo{(1d) while the two (2d) and three (3d) dimensional cases are investigated via direct numerical simulations. An example of a bifurcation diagram in 2d is given in \cite{EGUW2018springer}.} We expect such resting localized states (i.e., resting crystalline patches) to exist also for the aPFC model at least at small values of the activity parameter \lo{similar to the clusters observed at small activity in \cite{ReBH2013pre}.} Increasing activity brings the system further out of equilibrium and we expect that the localized states begin to travel. However, we also expect that activity might destroy the crystalline patches.

In general, localized states are experimentally observed and modeled in various areas of biology, chemistry and physics \cite{MathBio,BioPatterns,coulletPRL84,ChemWaves,BurkeKnoblochLSgenSHe}. Examples range from localized patches of vegetation patterns \cite{MERON2004}, local arrangements of free-surface spikes of magnetic fluids closely below the onset of the Rosenzweig instability \cite{richterPRL94} and localized spot patterns in nonlinear optical systems \cite{SchaepersPRL2000} to oscillating localized states (oscillons)  in vibrated layers of colloidal suspensions \cite{LiouPRL1999}.

In the context of solidification described by PFC models, localized states are observed in and near the thermodynamic coexistence region of liquid and crystal state.  Crystalline patches of various size and symmetry can coexist with a liquid environment depending on control parameters as mean density and undercooling \cite{RATK2012pre,TARG2013pre, EGUW2018springer}. For instance, increasing the mean density, the crystals are enlarged as further density peaks (or ``bumps'', or ``spots'') are added at their borders. Ultimately, the whole finite domain is filled and the branches of localized states terminate on the branch of space filling periodic states. Within their existence region, the localized states form ``snaking'' branches in the bifurcation diagram \cite{BurkeKnoblochSnakingChaos2007,SandstedeSnakes}. An important difference between conserved systems like the PFC model and non-conserved systems like the SH model, is that the respective snaking curves of localized states are slanted \cite{BoCR2008pre,Dawe2008sjads,LoBK2011jfm,PACF2017prf} and straight \cite{K_IMA16,BurkeKnoblochSnakingChaos2007,ALBK2010sjads,LSAC2008sjads}, respectively. For an extensive discussion of this point see the conclusion of Ref.~\cite{TARG2013pre}. \lo{Note that besides mass conservation also boundary conditions can have an influence on the type of snaking \cite{kozyreffPRL2009}.}

Here, we use the aPFC model to explore how slanted snaking of localized states as a characteristic feature of pattern-forming systems with a conserved quantity is amended by activity. This includes the question when and how resting localized states start to travel and whether and how they are destroyed by activity. Our work is organized as follows:
Section~\ref{sec:level1mod} introduces the model, its analytical and numerical treatment, while section~\ref{sec:level1lin} analyzes the linear stability of the uniform state (liquid state)
and discusses the different types of dispersion relations. Then, sections~\ref{sec:level1per} and~\ref{sec:level1loc} employ numerical continuation techniques to determine
bifurcation diagrams for resting and traveling periodic states (crystal) and localized states (crystallites coexisting with liquid), respectively, employing the mean density and activity parameter as main control parameters. Section~\ref{sec:level1drift} analyzes the condition for the onset of motion of crystallites. Finally, section~\ref{sec:level1con} concludes and gives an outlook.

\section{\label{sec:level1mod}The model}
\subsection{\label{sec:level2}Governing equations}
The local state variables of the aPFC model as introduced in Ref.~\cite{MenzelLoewen} are the scalar order parameter field $\psi(\mathbf{r},t)$, $\mathbf{r}\in \Omega \subset \mathbb{R}^\mathrm{n}$ (called in the following ``density'') where $\Omega$ denotes the considered domain, and the vectorial order parameter field $\mathbf{\ensuremath{P}}(\mathbf{r},t)$ (called in the following ``polar ordering'') that describes the local strength and direction of the active drive. The field $\psi(\mathbf{r},t)$ is conserved, i.e., $\int_\Omega\psi\, \mathrm{d^n r}$ is constant, and specifies the modulation about the mean density $\bar{\psi}$ that itself encodes the deviation from the critical point \cite{EmmerichPFC}. The field $\mathbf{\ensuremath{P}}(\mathbf{r},t)$ is non-conserved.

The uncoupled dynamics of $\psi(\mathbf{r},t)$ and $\mathbf{\ensuremath{P}}(\mathbf{r},t)$ corresponds to a purely conserved and a mixed non-conserved and conserved gradient dynamics on an underlying free energy functional $\mathcal{F}[\psi,\vec{P}]$, respectively. The functional contains no terms mixing the two fields and the coupling is purely non-variational, i.e., it can not be written as a gradient dynamics. The coupling is introduced in both equations in the simplest nontrivial form allowed for by the tensorial character of the fields that keeps the conserved character of the $\psi$-dynamics, i.e., the evolution of $\psi$ follows a continuity equation $\partial_t\psi=-\nabla\cdot\vec{j}$ where $\vec{j}$ is a flux. The non-dimensional evolution equations are \cite{MenzelLoewen}
\begin{align}
\partial_{t}\psi &=  \nabla^{2}\frac{\delta\mathcal{F}}{\delta\psi}-v_{0}\nabla\cdot\mathbf{P},\\
\partial_{t}\mathbf{P} &= \nabla^{2}\frac{\delta\mathcal{F}}{\delta\mathbf{P}}-D_{\mathrm{r}}\frac{\delta\mathcal{F}}{\delta\mathbf{P}}-v_{0}\nabla\psi
\label{eq:gov}
\end{align}
where $v_0$ is the coupling strength, also called activity parameter or velocity of self-propulsion. Physically speaking, $\mathbf{\ensuremath{P}}$ is subject to translational and rotational diffusion with $D_{\mathrm{r}}$ being the rotational diffusion constant. The functional $\mathcal{F}[\psi,\ensuremath{P}]$ is the sum of the standard phase-field-crystal functional $\mathcal{F}_{\mathrm{pfc}}[\psi]$ \cite{ElderGrantPRL88, ElderGrantPRE70, EmmerichPFC} and an orientational part $\mathcal{F}_{\mathbf{P}}[\vec{P}]$ 
\begin{equation}
\mathcal{F}=\mathcal{F}_{\mathrm{pfc}}+\mathcal{F}_{\mathbf{P}}
\end{equation}
with
\begin{equation}
 \mathcal{F}_{\mathrm{pfc}}[\psi] = \int \mathrm{d^nr}\left\{ \frac{1}{2}\psi\left[\epsilon+\left(1+\nabla^{2}\right)^{2}\right]\psi+\frac{1}{4}(\psi+\bar{\psi})^{4}\right\}
 \label{eq:functional}
\end{equation}
and 
\begin{equation}
\mathcal{F}_{\mathbf{P}}[\vec{P}]=\int \mathrm{d^nr} \left(\tfrac{C_1}{2}\mathbf{P}^{2}+\tfrac{C_2}{4}\mathbf{P}^{4}\right).
\label{eq:functionalP}
\end{equation}
The functional (\ref{eq:functional}) encodes the phase transition between liquid and crystal state \cite{EmmerichPFC}. It consists of a negative interfacial energy density ( $\sim|\nabla\psi|^2$)  that favors
the creation of interfaces, a bulk energy density and a stabilizing stiffness term ($\sim(\Delta\psi)^2$) -- this can be seen by partial integration. The parameter $\epsilon$ encodes temperature. Namely, negative values correspond to an undercooling of the liquid phase and result in solid (periodic) states for suitable mean densities $\bar{\psi}$, whereas positive values result in a liquid (homogeneous) phase. The functional (\ref{eq:functionalP}) with $C_1<0$ and $C_2>0$ allows for spontaneous polarization (pitchfork bifurcation at  $C_1=0$). However, in most of our work we will avoid spontaneous polarization using positive $C_1>0$ and $C_2=0$ as also done in most of the analysis of Refs.~\cite{MenzelLoewen,MenzelOhtaLoewenPhysRevE.89,ChGT2016el}. With $C_1>0$ diffusion reduces the polarization.

Determining the variations of Eqs.~(\ref{eq:functional}) and  (\ref{eq:functionalP}) and introducing them in the governing equations~(\ref{eq:gov}) we obtain the kinetic equations
\begin{align}
\partial_{t}\psi &= \nabla^{2}\left\{\left[\epsilon+\left(1+\nabla^{2}\right)^{2}\right]\psi+\left(\bar{\psi}+\psi\right)^{3}\right\}-v_{0}\nabla\mathbf{\cdot P}, \label{eq:dtpsi} \\
\partial_{t}\mathbf{P} &= C_1\nabla^{2}\mathbf{P} - D_{\mathrm{r}}C_1\mathbf{P}-v_{0}\nabla\psi. 
\label{eq:dtP}
\end{align}
In the following we study resting and traveling solutions of these equations in the spatially one-dimensional case with a special emphasis on the onset of motion.
Then $\vec{P}$ also becomes a scalar $P$ and indicates the strength and sense of direction of motion.

\subsection{\label{sec:level2}Steady and stationary states}
To investigate steady and stationary states  (where the latter are steady states in some comoving frame that moves with velocity $c$) we consider Eqs.~(\ref{eq:dtpsi}) and (\ref{eq:dtP}) with $\partial_{t}\psi=c\partial_x\psi$ and $\partial_{t}P=c\partial_x P$. Hence, positive velocities $c$ correspond to a propagation to the left. Then Eq.~(\ref{eq:dtpsi}) can be integrated once and we obtain the coupled fifth- and second-order ordinary differential equations 
\begin{align}
0 =& \partial_{x}\left\{\left[\epsilon+\left(1+\partial_{xx}\right)^{2}\right]\psi+\left(\bar{\psi}+\psi\right)^{3}\right\}-v_{0}P \label{eq:steadystatePSI} \nonumber \\
&-c \psi-J,\\
0 =& C_1 \partial_{xx}P-D_{\mathrm{r}}C_1 P - v_{0}\partial_{x}\psi-c\partial_{x}P \label{eq:steadystateP}
\end{align}
where the integration constant $J$ represents a flux. We emphasize that the velocity $c$ is equal to zero for resting states. For traveling states it is a nonlinear eigenvalue that has to be determined along with the solution profile. 

Beside the trivial steady state $(\psi=0, P=0)$ there exist spatially-modulated states $(\psi=\psi(x), P=P(x))$ that solve Eqs.~(\ref{eq:steadystatePSI}) and (\ref{eq:steadystateP}). We will determine their bifurcation diagrams employing continuation techniques (see next section). In the treated special case of $C_2=0$ [cf.~Eq.~(\ref{eq:dtP})], for periodic states one may integrate the linear Eq.~(\ref{eq:steadystateP}) over one period $\ell$ and finds $\int_\ell dx\,P(x)=0$. As $\int_\ell dx\,\psi(x)=0$ by definition, Eq.~(\ref{eq:steadystatePSI}) then implies $J=0$. Note, that as $\psi(x)$ is the deviation from the mean $\bar\psi$, for $J=0$ the flux of material is given by $c\bar\psi$. Note, that the system is invariant under the transformation $(\psi,P,x,c)\to (\psi,-P,-x,-c)$. In the case of $\bar\psi=0$, also the symmetry $(\psi,P,x,c)\to (-\psi,-P,x,c)$ holds.

\subsection{\label{sec:level2}Numerical approach}
We employ numerical path-continuation techniques \cite{DWCD2014ccp,KrauskopfOsingaGalan-Vioque2007,Kuznetsov2010,EGUW2018springer} bundled in the package auto07p \cite{DoKK1991ijbc,Doedelauto2012} to determine steady ($c=0$) and stationary ($c\neq0$) periodic and localized solutions of Eqs.~(\ref{eq:steadystatePSI}) and (\ref{eq:steadystateP}) on a domain of size $L$. The techniques allow one to follow branches of solutions in parameter space, detect bifurcations, switch branches and in turn follow the bifurcating branches. The pseudo-arclength continuation implemented in auto07p is also able to follow branches when they fold back at saddle-node bifurcations allowing one to determine the entire bifurcation diagram. In the literature the method is extensively applied to the SH equation \cite{BuKn2006pre,MaBK2010pd,BuDa2012sjads} and PFC-type models \cite{Thie2010jpcm,TARG2013pre,RATK2012pre}. To our knowledge, continuation has not yet been applied to the aPFC model.

To do so, our system of Eqs. (\ref{eq:steadystatePSI}) and (\ref{eq:steadystateP}) is transformed into a seven-dimensional dynamical system (with $x$ being the independent variable with seven periodic boundary conditions). A phase condition that breaks translational invariance and a constraint that controls the volume are included as integral conditions (cf.~Refs.~\cite{cenosTutorial, EGUW2018springer} for examples of using such conditions for several related equations). This implies that in each continuation run beside the main control
parameter one has two further parameters that have to be adapted (with other words they represent nonlinear eigenvalues of the problem). Here, we use either the mean density $\bar{\psi}$ or the activity $v_{0}$ as main control parameter while velocity $c$ and flux $J$ are adapted. 

The resulting bifurcation diagrams are given in terms of the $L^2$-norm of the solution array that we use as main solution measure. 
It is defined by
\begin{equation}
||\underline{\psi},\underline{P}||_2=\sqrt{\frac{1}{L} \int_0^L \sum_{i=1}^7 a_i^2 \mathrm{d}x}
\end{equation}
where the $a_i$ stand for the elements of the solution array $(\underline{\psi},\underline{P})=(\psi, \partial_{x}\psi, \partial^2_{x}\psi, \partial^3_{x}\psi, \partial^4_{x}\psi, P, \partial_{x} P)$.

In addition, we perform direct numerical simulations (DNS) employing a pseudo-spectral method. Starting from a homogeneous state with a small random perturbation, Eqs.~(\ref{eq:dtpsi}) and (\ref{eq:dtP}) are integrated forward in time via a semi-implicit Euler method, while spatial derivatives are calculated in Fourier space and nonlinearities in real space. 
\section{\label{sec:level1lin}Liquid state and its linear stability}
\begin{figure*}%[!hbt]
\includegraphics{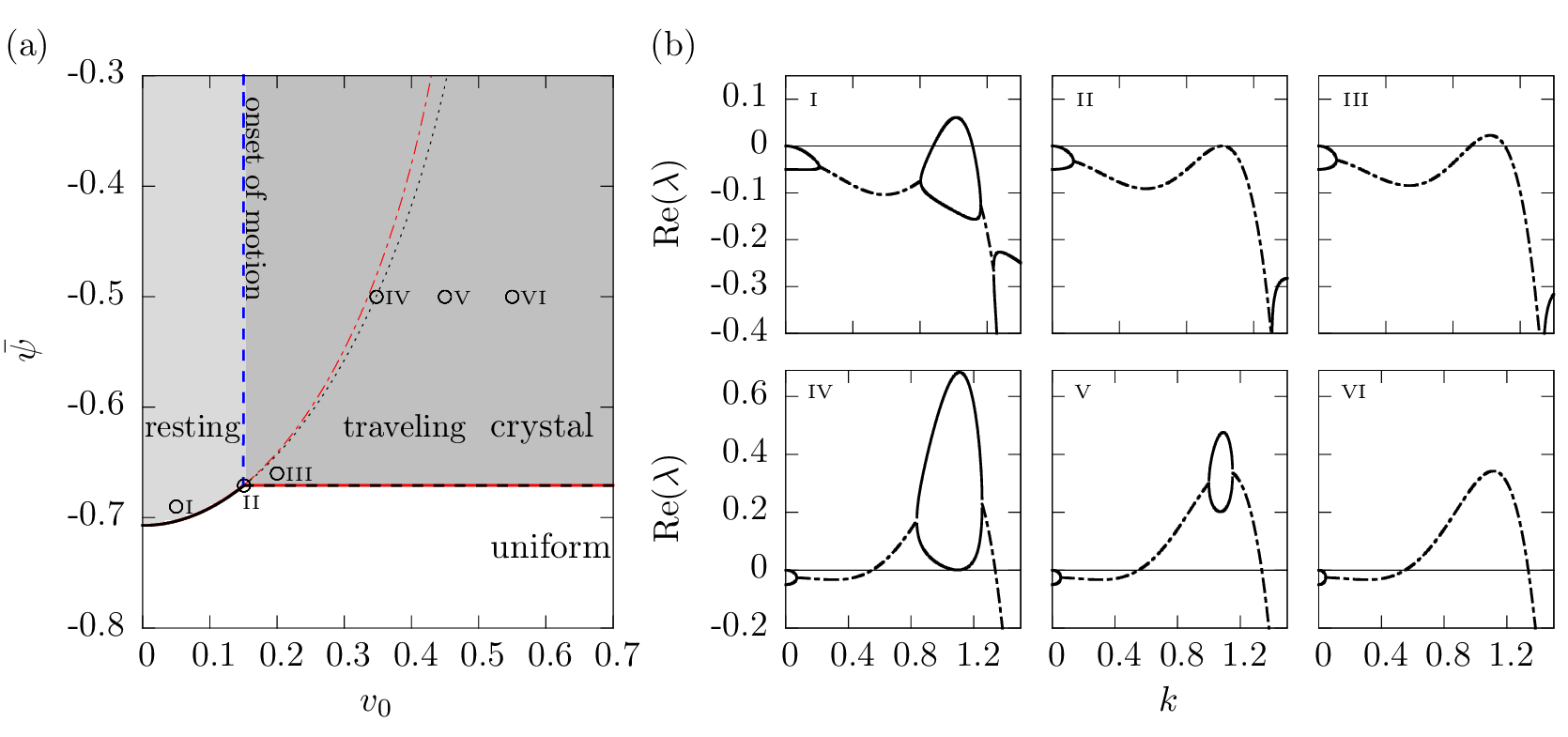}
\caption{\label{fig:phasediagram} \lo{(a) Morphological phase diagram of the active PFC model in the 1d case in the plane spanned by activity $v_{0}$ and mean concentration $\bar{\psi}$ as obtained by linear and nonlinear analysis. The remaining parameters are $\epsilon=-1.5$, $C_{1}=0.1$, $C_{2}=0.0$ and $D_\mathrm{r}=0.5$.  Labels ``I'' to ``VI'' in (a) indicate parameters for which the real part of the dispersion relation $\lambda(k)$ is shown in (b) with solid [dashed] lines corresponding to real [complex] eigenvalues. In (a) gray shading indicates the linearly unstable region where $\mathrm{Re}(\lambda(k))>0$ for a band of wavenumbers $k$. There, periodic (crystalline) patterns are formed. The analytically obtained curved solid and horizontal dashed black lines indicate the onset of the monotonic and oscillatory finite wavelength instability, respectively. For the coinciding red lines the critical wavenumber is approximated as $k_c\approx 1$. The gray-shaded region of the linearly unstable homogeneous (liquid) phase is separated by the vertical dashed blue line into regions where stable resting (light gray) and stable traveling (dark gray) crystals are found in the fully nonlinear regime. The thin dotted black and dot-dashed red lines indicate changes in the primary bifurcation behavior and indicate where the (then unstable) resting crystals cease to exist (see main text).}}
\end{figure*}

The trivial solution of a PFC model is the homogeneous state that represents the liquid phase where on diffusive time scales the probability of finding a particle is uniform in space. In analogy, we also call the 
homogeneous state $(\psi_0,\mathbf{P}_{0})=(0,\mathbf{0})$ of the present aPFC model ``liquid phase''. Although it exists at all parameter values, for $\epsilon<0$ it is only stable at high $|\bar{\psi}|$ and at lower $|\bar{\psi}|$ becomes unstable w.r.t. coupled density and polarization fluctuations. However, in the context of colloidal particles the region $\bar{\psi}>0$ is unphysical \cite{TARG2013pre} and we focus on $\bar{\psi}<0$ where the liquid state is stable at low values of $\bar{\psi}$ (high $|\bar{\psi}|$) while the crystalline state is at high $\bar{\psi}$ (low $|\bar{\psi}|$).
To determine the linear stability of the homogeneous state, Eqs.~(\ref{eq:dtpsi}) and (\ref{eq:dtP}) are linearized in small perturbations $(\delta\psi,\delta\mathbf{P})$ about \lo{$(0,\mathbf{0})$} yielding 
\begin{align}
\partial_{t}\delta\psi &= \nabla^{2}\left(\epsilon+3\bar{\psi}^{2}+\left(1+\nabla^{2}\right)^{2}\right)\delta\psi-v_{0}\nabla \cdot \delta\mathbf{P}, \label{eq:dtdeltaPSI_trivial}\\
\partial_{t}\delta\mathbf{P} &= \nabla^{2}\left(C_{1}\delta\mathbf{P}\right)-D_{\mathrm{r}}C_{1}\delta\mathbf{P}-v_{0}\nabla\delta\psi \label{eq:dtdeltaP_trivial}.
\end{align}
We restrict our analysis to one spatial dimension, expand the spatial dependency of the perturbation into decoupled harmonic modes and, in consequence, use the exponential ansatz $\delta \psi (x,t), \delta P(x,t) \propto \mathrm{ext}(ikx+\lambda t)$ 
in Eqs.~(\ref{eq:dtdeltaPSI_trivial}) and (\ref{eq:dtdeltaP_trivial}) to obtain the eigenvalues
\begin{equation}
\lambda_\pm = \frac{1}{2}\left(L_1(k)+L_2(k)\right)\pm \frac{1}{2} \sqrt{\left(L_1(k)-L_2(k)\right)^2-4 v_0^2 k^2} \label{eq:eigenvalues}
\end{equation}
where 
\begin{align}
 L_1(k)&= -k^2\left( \epsilon + 3 \bar{\psi}^2 + \left( 1-k^2 \right)^2 \right)\\
 L_2(k)&= -k^2 C_1 -D_\mathrm{r}C_1.
\end{align}
We investigate the stability of \lo{$(\psi_0,P_{0})=(0,0)$} in the $(\bar{\psi},v_{0})$-plane and determine the boundary, where the largest real part of an eigenvalue $\lambda$ crosses zero at a finite critical wavenumber $k_c$, i.e., a maximum of the dispersion relation $\mathrm{Re}(\lambda(k))$ touches zero. This can either occur with a zero or with a finite imaginary part corresponding to unstable modes that result in the development of a resting or traveling crystalline state (i.e., spatially-periodic solution), respectively. Setting $\lambda=0$ and substituting $k^{2}=z$ gives a cubic equation for $z$. Considering Cardano's method and the desired number of roots, we are able to find analytical expressions for the stability boundaries in both cases.

The results are presented in Fig.~\ref{fig:phasediagram}(a). The white area at low $\bar{\psi}$ corresponds to a linearly stable liquid phase, whereas the gray shading marks regions where the liquid phase is unstable w.r.t.\ spatially periodic perturbations. The dashed horizontal line (red and black) \lo{at $\bar{\psi}\approx-0.67$} separates the linearly stable liquid phase and a traveling crystal. It is independent of activity $v_0$, as can be seen, when taking a closer look at Eq.~(\ref{eq:eigenvalues}). There, $v_0$ only appears in the (then negative) discriminant and therefore only influences $\mathrm{Im}(\lambda)$, i.e., the drift velocity $c$ of the perturbation modes. The upwards curved black line that separates white and light gray regions at low activity indicates the stability border of the liquid phase related to a purely real eigenvalue, i.e., a monotonic instability.
Alternatively to Cardano's method, the critical wavenumber can be approximated by $k_c\approx 1$ as used in Ref.~\cite{ChGT2016el}. This approximation gives the red lines in Fig.~\ref{fig:phasediagram}(a). The resulting stability border can not be distinguished by eye from the exact results.

\begin{figure}
 \includegraphics{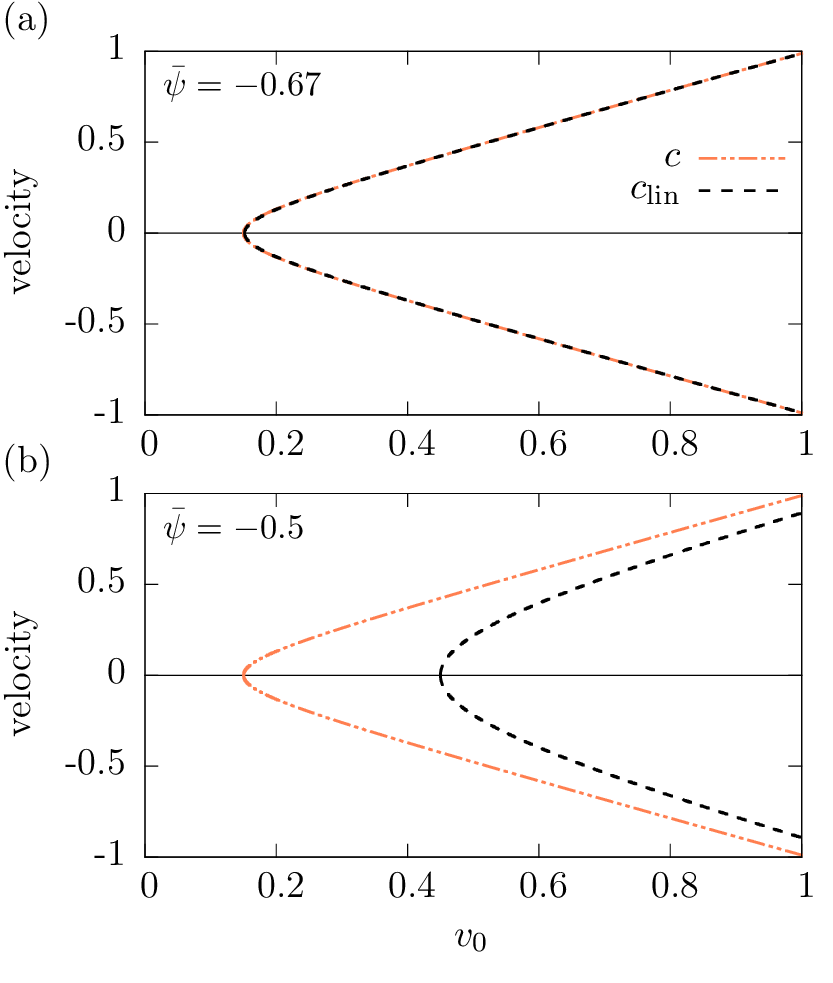}
 \caption{ \label{fig:im_vs_c} \lo{Velocity $c_\mathrm{lin}$ of the dominant linear mode (black dashed line) and drift velocity $c$ (orange dot-dashed) of fully nonlinear moving crystals in dependence of activity $v_0$ for (a) $\bar{\psi}=-0.67$ and (b) $\bar{\psi}=-0.5$. Remaining parameters as in Fig.~\ref{fig:phasediagram}. The eigenvalues are obtained from the linear stability analysis of the homogeneous state. The velocity $c$ of the fully nonlinear traveling crystals is determined by numerical continuation. The speed of linear modes $c_\mathrm{lin}$ corresponds to Im$(\lambda)/k$, i.e., $c_\mathrm{lin} =\mathrm{Im}(\lambda)$ for $k=1$.
(a) In the linear regime close to the onset of crystallization $c_\mathrm{lin}$ and $c$ coincide. (b) Deep in the unstable regime, $c_\mathrm{lin}$ does not provide a suitable approximation for the onset of motion of the crystal. However, $c_\mathrm{lin}$ and $c$ approach each other at high $v_0$.}}
\end{figure}

Corresponding dispersion relations are displayed in Fig.~\ref{fig:phasediagram}(b) showing $\mathrm{Re(\lambda)}$ of the \lo{leading two eigenvalues} with solid [dashed] lines for real [complex] eigenvalues. The roman numbering corresponds to labels in the stability diagram~\ref{fig:phasediagram}(a). \lo{Case~I shows a dominant (i.e.~at the maximum) instability mode that is real (i.e.~monotonic), and likely results in a resting crystal.} However, with increasing activity $v_0$ the 'bubble' of real eigenvalues around the maximum shrinks. At the codimension-2 point (case II) this bubble shrinks to zero and the \lo{marginally stable eigenvalue at the maximum becomes complex. Case~III then shows a dominant mode  that is complex (i.e.~oscillatory), and likely results in crystallization into a traveling crystal. Cases~IV to VI give further qualitatively different dispersion relations. In particular, points~V and VI illustrate the important change in the character of the dominant mode at $k\approx1$ from monotonic to oscillatory. Case~IV is located on the thin dotted black line in Fig.~\ref{fig:phasediagram}(a) that marks where the minimum of Re($\lambda$) touches zero. \lo{The dot-dashed red line is the corresponding approximation obtained by assuming $k_\mathrm{min}=1$.
Crossing this line does not influence the linear stability but changes the number of expected primary bifurcations. Accordingly, in Fig.~\ref{fig:crystal_v0} below (that represents a horizontal cut through Fig.~\ref{fig:phasediagram}(a) at $\bar{\psi}=-0.5$), at $v_0\approx0.34$ the branch of the (then unstable) resting crystals ends in a subcritical bifurcation.}

As discussed above the two phase boundaries in Fig.~\ref{fig:phasediagram}(a) between the liquid phase and, respectively, \lo{stable resting and stable traveling crystals} collide in point II. From there, the boundary between fully nonlinear resting and traveling crystals continues nearly vertically upwards (blue dashed line). In the nonlinear regime, this separating line cannot be determined by the present linear considerations and is obtained by numerical continuation. The resulting dashed blue line marks the onset of crystal motion and confirms Ref.~\cite{ChGT2016el}, where a similar straight line in a different parameter plane was deduced from direct time simulations.  \lo{Note that to the right of the vertical line, there is the region where resting crystals still exist as unstable steady states.}

\lo{Comparing the velocity $c_\mathrm{lin}$ of the dominant linear mode and the fully nonlinearly determined drift velocity $c$ allows us to rate how well the linear analysis performs. Fig.~\ref{fig:im_vs_c}(a) shows that close to but above the liquid-solid boundary at $\bar{\psi}=-0.67$ (Fig.~\ref{fig:phasediagram}), the linear (dashed black line) and the fully nonlinear results  (dot-dashed orange line) coincide in the onset of motion and the drift velocity in the entire $v_0-$range. However, in the nonlinear regime at $\bar{\psi}=-0.5$, Fig.~\ref{fig:im_vs_c}(b) shows that there is a considerable offset in the onset of motion. Yet, at high activities $v_0$ the linear and nonlinear velocities still converge. The nonlinear drift velocity $c$ corresponds to the branch of traveling crystals shown in Fig.~\ref{fig:crystal_v0} in the next section.}
}

\section{Crystalline states}
\label{sec:level1per}
\begin{figure}
\includegraphics{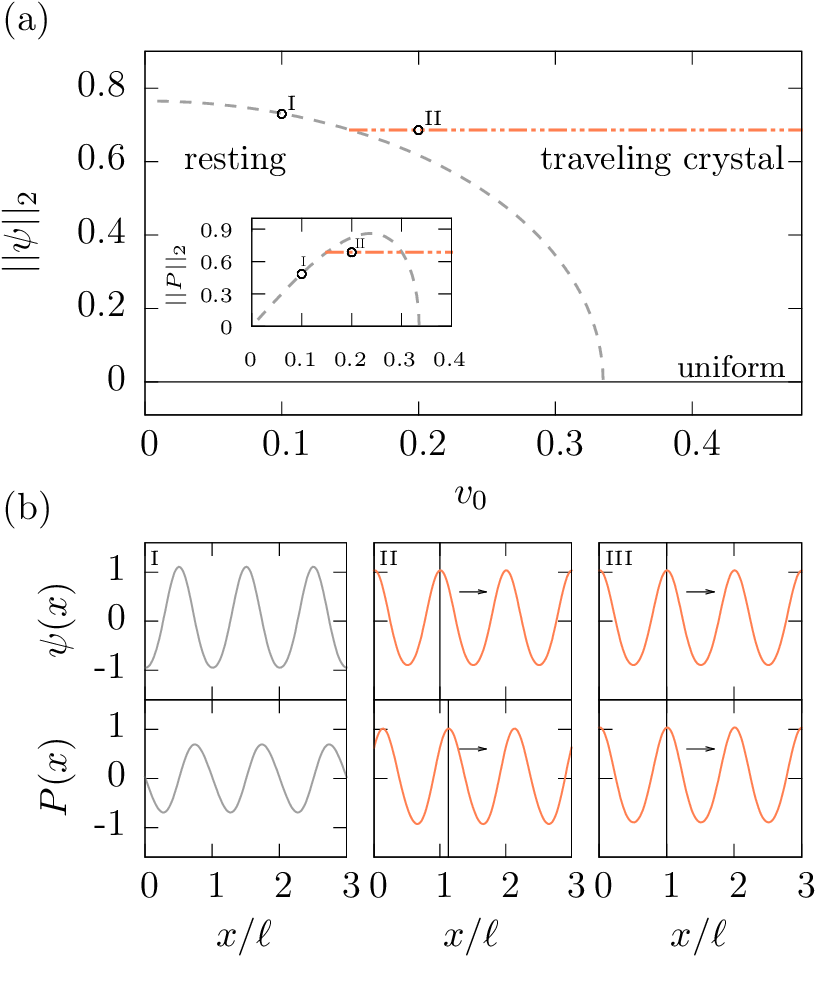}
\caption{\label{fig:crystal_v0}Resting and traveling crystals as a function of activity $v_0$ in the one-dimensional aPFC model. (a) The solution profiles of the periodic crystalline states are characterized by the $L^2$-norms of $\psi$, \lo{$||\psi||_2=\sqrt{\frac{1}{L} \int_0^L \psi^2 \mathrm{d}x}$}, and $P$ (inset). Branches of resting structures are shown in dashed gray, while traveling crystals are in dot-dashed orange. At a critical value of $v_0\approx0.15$, the resting crystal is destabilized and starts to move. The spatial periodicity remains unchanged.  (b) depicts parts (3 times period $\ell$) of the profiles of the structures at points indicated by roman numbers in (a). Crystals I and II are close to the onset of motion. Profile III shows an active crystal at a high activity of $v_0=10.0$ beyond the range of (a). The drift velocity $c$ of the moving crystals increases monotonically with $v_0$ as \lo{shown in Fig.~\ref{fig:im_vs_c}(b). Note that the phase difference} between $\psi$ and $P$ changes when varying $v_0$, highlighted by vertical lines. $\bar{\psi}=-0.5, L=100$, remaining parameters are as in Fig.~\ref{fig:phasediagram}.}
\end{figure}

In the standard PFC model (Eq.~(\ref{eq:steadystatePSI}) with $v_0=0$), at sufficient distance from the critical point ($\epsilon$ sufficiently negative or $|\bar{\psi}|$ sufficiently low) the transition from the liquid state (homogeneous solution) to a crystalline state (periodic solution) corresponds to a first order liquid-solid phase transition with a parameter region - limited by the binodal lines - where the two states coexist \cite{TARG2013pre}. As $\psi$ is a conserved quantity this does not automatically imply that one has a subcritical bifurcation from the homogeneous to the periodic solution branch. For a detailed discussion of this intricate point see the conclusion of Ref.~\cite{TARG2013pre}. 

Here, as the aPFC model is non-variational the transition between the states does not correspond anymore to a thermodynamic phase transition, i.e., arguments based on free energy do not hold anymore.  Furthermore, now also the activity $v_0$ may be used to induce the transition. In particular, for the parameters of Fig.~\ref{fig:phasediagram} at $\bar{\psi}$ approximately between $-0.71$ and $-0.67$ increasing $v_0$ beyond the solid line melts the resting crystal.
More striking is the behavior at higher densities (in Fig.~\ref{fig:phasediagram}(a)) for  $\bar{\psi}$ above $-0.67$). As illustrated in the bifurcation diagram Fig.~\ref{fig:crystal_v0}, there, increasing  $v_0$ does not destroy the resting crystal but results in the onset of motion at a critical activity $v_c\approx0.15$ (corresponding to the vertical dashed line in Fig.~\ref{fig:phasediagram}(a)), \lo{i.e., in a transition from a stable resting to a stable traveling crystal.}

Specifically, for the resting crystals Fig.~\ref{fig:crystal_v0}(a) shows that with increasing activity the norm of $\psi$ monotonically decreases while, in contrast, the amplitude of the polarization field (see inset) first increases from zero (at $v_0=0$) until at some $v_0=v_c$ its norm equals the one of $\psi$. There the branch of traveling crystals bifurcates and the resting crystals become unstable and ultimately cease to exist \lo{(after further undergoing a Hopf bifurcation)} at about $v_0=0.34$ where the branch ends in a subcritical pitchfork bifurcation on the branch of homogeneous states. \lo{As mentioned in section~\ref{sec:level1lin}, this bifurcation corresponds to point~IV in Fig.~\ref{fig:phasediagram}. There, a double real eigenvalue of the linear stability problem of the liquid state crosses zero indicating a bifurcation of the uniform state. The mentioned unstable steady and oscillatory states will be discussed elsewhere.}
% to remember: in periodic setting cos and sin modes cross zero, therefore 2 EV become unstable simultaneously and two phase-shifted branches emerge

At $v_c$, a drift-pitchfork bifurcation \cite{Friedrich2005} occurs, i.e., a real eigenvalue crosses zero (see stability analysis in section~\ref{subsec:linstab}) and two branches of moving periodic states (i.e., traveling crystals) emerge from the branch of resting crystals. An analytical condition for the drift bifurcations is derived in section~\ref{sec:level1drift}.
The two bifurcating branches with the same norm are related by the symmetry  $(\psi,P,x,c)\to (\psi,-P,-x,-c)$) and the velocity close to the bifurcation is $c\propto(v_0-v_c)^{1/2}$. The individual solutions on the emerging branches do not have the symmetry $(\psi,P,x)\to (\psi,-P,-x)$ anymore that the resting crystal states have (i.e., zero crossings of $P$ do not anymore coincide with the position of the peak maxima of $\psi$). Instead, for the traveling crystals the individually practically unchanged $\psi(x)$ and $P(x)$ profiles are shifted w.r.t.\ each other. The profiles keep their spatial periodicity and always move with a constant drift velocity. This velocity and the size of the phase shift between $\psi$ and $P$ profiles increase monotonically with $v_0>v_c$ also far away from the bifurcation. Indeed, for $v_0\gg1$ one finds $c\approx v_0$ and $\psi(x) \approx P(x)$. Typical density and polarization profiles are given in Fig.~\ref{fig:crystal_v0}(b).
\section{\label{sec:level1loc}Localized states}

As for the passive PFC model, where the described phase transition between liquid and crystal state is of first order for sufficiently negative $\epsilon$, one finds that in the transition region patches of liquid state and crystal state may coexist. In the PFC model this corresponds to the existence of a broad variety of spatially localized states (or crystallites) that in 1d were numerically analyzed in Ref.~\cite{RATK2012pre,TARG2013pre} (for selected 2d results see \cite{EGUW2018springer}). Next we systematically explore how the bifurcation structure of these crystallites is amended by activity employing Eqs.~(\ref{eq:steadystatePSI}) and (\ref{eq:steadystateP}). We investigate if and to what extent the phenomenon of slanted homoclinic snaking \cite{SandstedeSnakes} is changed by finite values of activity. Do traveling localized states arise due to self-propulsion? Can motion also be induced by changes in the mean concentration? 

Following former works, we classify the localized states according to their spatial symmetry \cite{BurkeKnoblochLSgenSHe,TARG2013pre} and their drift velocity \cite{MenzelLoewen}. There are two kinds of resting localized states (RLS) that have a parity (left-right) symmetry in the $\psi$-component and an inversion-symmetric polarization: $(\psi(x),P(x))=(\psi(-x),-P(-x))$. The symmetric localized patches can either have a $\psi$-peak or $\psi$-trough at the center, resulting in an odd or even number of peaks, respectively. We call them ``odd states'' ($\mathrm{RLS_{\mathrm{odd}}}$) and ``even states'' ($\mathrm{RLS_{\mathrm{even}}}$). Beside spatially symmetric states, resting asymmetric localized states exist that have no parity symmetry. We refer to them as $\mathrm{RLS_{\mathrm{asym}}}$. In the PFC model, the RLS states form an intricate tilted snakes-and-ladders structure \cite{TARG2013pre}.
Traveling localized states have a nonzero drift velocity and are called TLS. For TLS, the above symmetries in $\psi$ and $P$ are not preserved. 
 
\begin{figure}
\includegraphics{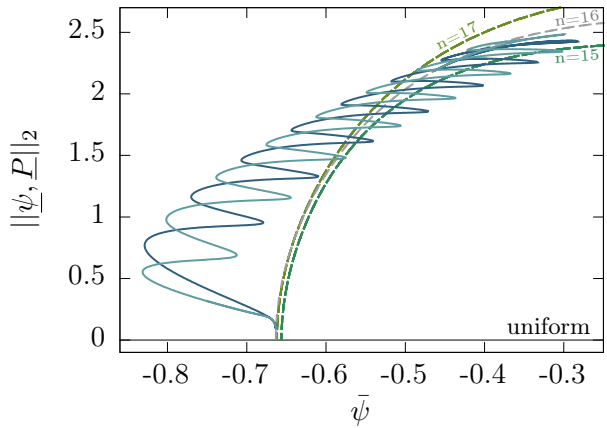}
\caption{\label{fig:snaking} Slanted homoclinic snaking of resting symmetric steady states (drift velocity $c=0$). Shown is the $L^2$-norm of the steady states  in dependence of the mean concentration $\bar{\psi}$. The active drive is fixed at $v_0=0.16475$. The steady localized states bifurcate subcritically from the periodic solution with $n=16$ peaks (dashed gray line). The light (dark) blue line represents resting localized structures with a peak (trough) in the middle, $\mathrm{RLS_{\mathrm{odd}}}$ ($\mathrm{RLS_{\mathrm{even}}}$). Both lines ultimately terminate on the $n=16$ periodic state. Beside the spatially extended crystal with $n=16$ peaks, there are solutions with $n=15$ and $n=17$ peaks (dashed green lines). Remaining parameters as in Fig.~\ref{fig:crystal_v0}.}
\end{figure}

\begin{figure}
\includegraphics{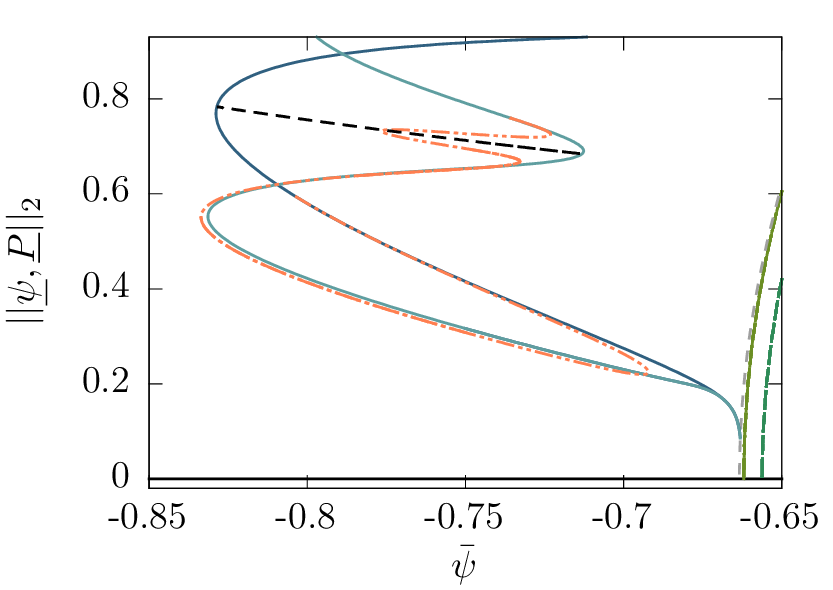}
\caption{\label{fig:firstladder}Resting and traveling localized states as a function of the mean concentration $\bar{\psi}$. The localized states are created in a subcritical bifurcation and branch off from the $n=16$ periodic solution branch (dashed gray, more periodic branches in dashed green). Light and dark blue lines: $\mathrm{RLS_{\mathrm{odd}}}$ and $\mathrm{RLS_{\mathrm{even}}}$. The ladder branch (dashed black line) corresponding to asymmetric states connects the two symmetric RLS. Beside snaking branches and the ladder rungs, we find traveling localized states (TLS, dot-dashed orange) that arise due to activity. Remaining parameters as in Fig.~\ref{fig:snaking}.}
\end{figure}

\begin{figure}
\includegraphics{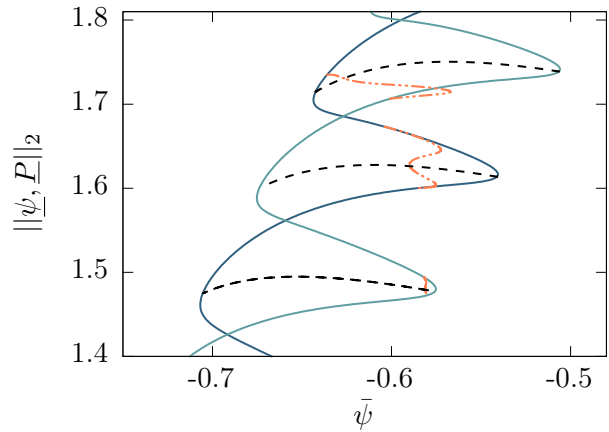}
\caption{\label{fig:snake-ladder}Tilted snakes-and-ladders structure for finite active drive. The light (dark) blue line represents odd (even) symmetric localized structures. The dashed black lines correspond to asymmetric localized states. Because of the active drive above $v_c$ there exist traveling states (TLS, dot-dashed orange line) that emerge in various drift bifurcations. The shown branches of TLS have between 5, 6 and 7 peaks in $\psi$. Remaining parameters as in Fig.~\ref{fig:snaking}.}
\end{figure}

\begin{figure}
\includegraphics{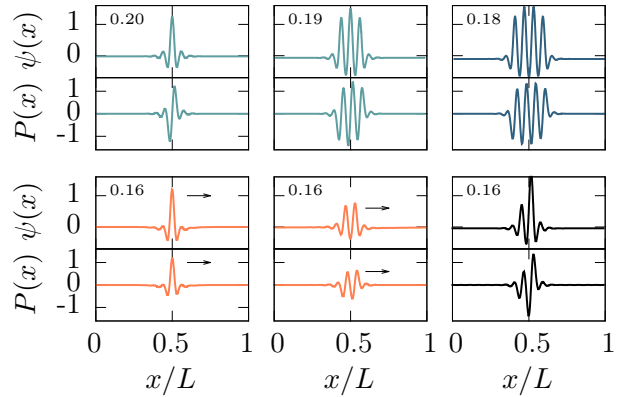}
\caption{\label{fig:solutions}Typical density and polarization profiles of localized states for $\bar{\psi}=-0.75$ and various values of activity $v_0$ (rounded value given in each panel). Blue colors indicate symmetric RLS. Two states with an odd number of peaks are followed by an even RLS (top, from left to right). An asymmetric resting state is plotted in black. The profiles in red are traveling localized states. Their profile is slightly asymmetric, too. Note that the integral over $\psi$ vanishes, as it only describes the modulation around $\bar{\psi}$. Remaining parameters as in Fig.~\ref{fig:crystal_v0}.}
\end{figure}

\subsection{Bifurcation diagrams}
Figure \ref{fig:snaking} gives the bifurcation diagram for periodic and localized states of the aPFC model for fixed finite activity $v_0\approx0.16>v_c$ employing the mean density $\bar\psi$ as control parameter. It illustrates the main characteristics of the resting crystallites (steady localized states) and their snaking path towards a spatially extended crystal that fills the whole considered domain. The appearance of the bifurcation diagram is similar to the one obtained for the conserved Swift-Hohenberg equation (passive PFC) \cite{TARG2013pre}, note, in particular, the slanted snaking that also occurs for other systems with conserved quantities \cite{BoCR2008pre,Dawe2008sjads,LoBK2011jfm,TARG2013pre}. The liquid state with solution measure $||\underline{\psi},\underline{P}||_2=0$ is destabilized when $\bar{\psi}$ is increased above a critical mean concentration $\bar\psi_c\approx-0.66$\lo{, coinciding with point II in Fig.~\ref{fig:phasediagram}(a)}. For the employed domain size of $L=100$, three branches of periodic states bifurcate from the uniform state. The dashed gray and dashed green lines correspond to periodic structures with $n=15,16$ and 17 $\psi$-peaks. Slightly beyond the primary bifurcation, the periodic state with $n=16$ is destabilized and two branches (light and dark blue) emerge in a subcritical secondary bifurcation. Fig.~\ref{fig:firstladder} gives a zoom of this region. The two branches correspond to the resting odd and even localized states, respectively. Both branches undergo a series of saddle-node bifurcations where their stabilities change (cf.~Fig.~\ref{fig:snake-ladder} and subsection~\ref{subsec:linstab}). The odd and the even branch of symmetric RLS are connected by many branches of asymmetric RLS that are given in Figs.~\ref{fig:firstladder} and \ref{fig:snake-ladder} as dashed black lines, but are not included in Fig.~\ref{fig:snaking}.

Each pair of saddle-node bifurcations adds a couple of peaks to the localized crystalline patch that, in consequence, enlarges until ultimately the whole domain is filled with the crystalline state and the branches of localized states terminate on the $n=16$ branch of periodic states.
Because of the conserved character of the density $\psi$ the density of the coexisting uniform state is not constant but changes with the increasing size of the crystalline patch. This results in the slanted snaking structure, i.e., the loci of subsequent saddle-node bifurcations do not form straight vertical lines in Fig.~\ref{fig:snaking} but drift towards larger $\bar\psi$. Increasing the domain size, adds more 'undulations' to the slanted snaking structure and the relative tilt between subsequent saddle-node bifurcations becomes smaller, however, without changing the overall tiltedness.

A qualitatively new feature of the solution structure of the aPFC model are the branches of traveling localized states (TLS) shown as dot-dashed orange lines in Figs.~\ref{fig:firstladder} and \ref{fig:snake-ladder}. The TLS drift with a constant velocity $c$. Their $\psi$ profiles look quite similar to the one of RLS, the left-right symmetry is broken, though. Crossing the onset of motion, the $P$ profile loses its inversion symmetry and approaches the phase and shape of $\psi$. Typical profiles of RLS and TLS are presented in Fig.~\ref{fig:solutions}. The branches of TLS bifurcate in drift-transcritical bifurcations from the branches of asymmetric RLS and in drift-pitchfork bifurcations from the branches of symmetric RLS. An analytical condition for the detection of the drift bifurcations is derived in section~\ref{sec:level1drift}. This criterion holds for both types of drift bifurcations.

The branches of TLS connect the snaking branches of symmetric RLS like rungs. They may connect two sub-branches of the same symmetry like the two lower orange branches in Fig.~\ref{fig:snake-ladder} as well as branches of $\mathrm{RLS_{\mathrm{odd}}}$ and $\mathrm{RLS_{\mathrm{even}}}$ like the orange branch with the highest norm in Fig.~\ref{fig:snake-ladder}. TLS of small extension (one or two peaks, i.e., the ones in Fig.~\ref{fig:firstladder}) exist in a broad range of mean density $\bar{\psi}$. Because of their similar profiles, the norm of RLS and TLS is almost equal and the branches seem to nearly coincide in the lower part of Fig.~\ref{fig:firstladder}.

\begin{figure*}
\includegraphics{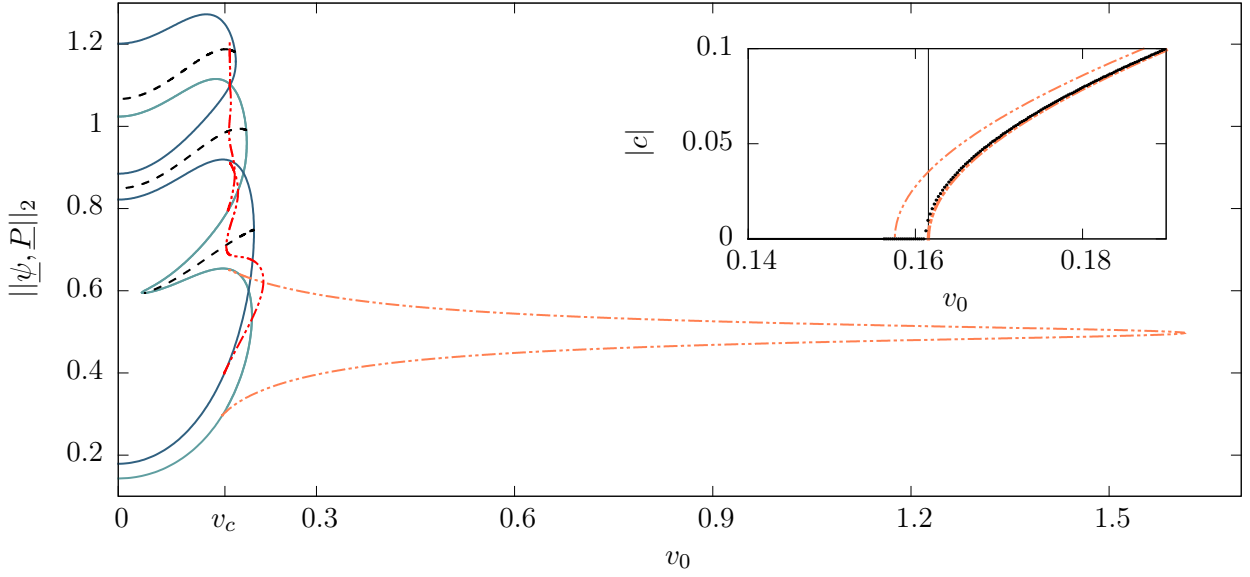}
\caption{\label{fig:bifv0} Bifurcation diagram of resting and traveling localized states giving the
$L^2$-norm as a function of the active drive $v_0$. The mean concentration is fixed at $\bar{\psi}=-0.75$. Resting solutions are indicated by blue (left-right symmetric states) and dashed black (asymmetric states) lines, moving states are dot-dashed red and orange. The traveling single peak exists up to high values of $v_0\approx1.6$. The remaining parameters are $\epsilon=-1.5$, $C_1=0.1$ and $D_\mathrm{r}=0.5$. Inset: Velocity $|c|$ of the traveling single peak as a function of $v_0$. At a critical value of $v_0=v_c$ (vertical line) the transition from a resting to a traveling linearly stable state occurs. Black dots (red dashed lines) give the results of direct numerical simulations (numerical continuation). The moving state corresponds to the long finger in the large panel. Its upper half is stable (right orange branch in inset). $v_c$ in the main plot and the vertical black line in the inset mark the onset of motion as calculated semi-analytically for the single peak (cf.~section \ref{sec:level1drift}).}
\end{figure*}

Similar to the case of periodic states, also for RLS an increase of the activity $v_0$ at fixed $\bar\psi$ may result in a transition to TLS. Fig.~\ref{fig:bifv0} gives a typical example of a bifurcation diagram using $\bar{\psi}=-0.75$. Thereby, the threshold value for the onset of motion slightly differs for the various RLS (inset of Fig.~\ref{fig:bifv0}). All discussed TLS have density and polarization profiles that are steady in corresponding comoving frames. 

Recall that the onset of motion coincides with a symmetry breaking related to a phase shift between the density and the polar ordering profiles. The density peaks are shifted away from the zeros of $P$, resulting in a nonzero value when integrating $\psi$ times $P$ over the width of a peak. Above the critical activity the left-right symmetry of the density profile is also broken. The same holds for the inversion symmetry of the polarization. As described above and shown in Fig.~\ref{fig:solutions} at large $v_0$ the $P$ profile approaches the position and shape of $\psi$. In fact, the norm of $\psi$ and $P$ are equal for traveling structures. 

Beside path-continuation we also employ direct time simulations of Eqs.~(\ref{eq:dtpsi}) and (\ref{eq:dtP}) to investigate the TLS. In particular, we track the traveling single density peak over time and determine its velocity. This confirms the continuation results  as shown in the inset of Fig.~\ref{fig:bifv0}. The two orange dot-dashed lines in the inset correspond to the long nose of a traveling single peak in the main panel. The upper branch of this nose is stable, losing its stability at the fold at $v_0 \approx 1.6$. The lower branch is unstable and corresponds to the left orange branch in the inset. Its onset of motion is at a slightly smaller value of $v_0$ as compared to the stable one. For the particular value of mean concentration $\bar{\psi}$ shown in Fig.~\ref{fig:bifv0}, localized states consisting of more than one peak appear to only exist in a fairly narrow range of $v_0$ around $v_c$. The dot-dashed red line in Fig.~\ref{fig:bifv0} that corresponds to broader TLS with a few peaks wiggles about an almost vertical line before terminating on the blue branch of four connected resting peaks. The region of existence of the TLS is studied via fold continuation in the next section. Note that the velocities of all these different traveling structures are very similar. 

\subsection{Fold continuation}

\begin{figure*}
\includegraphics{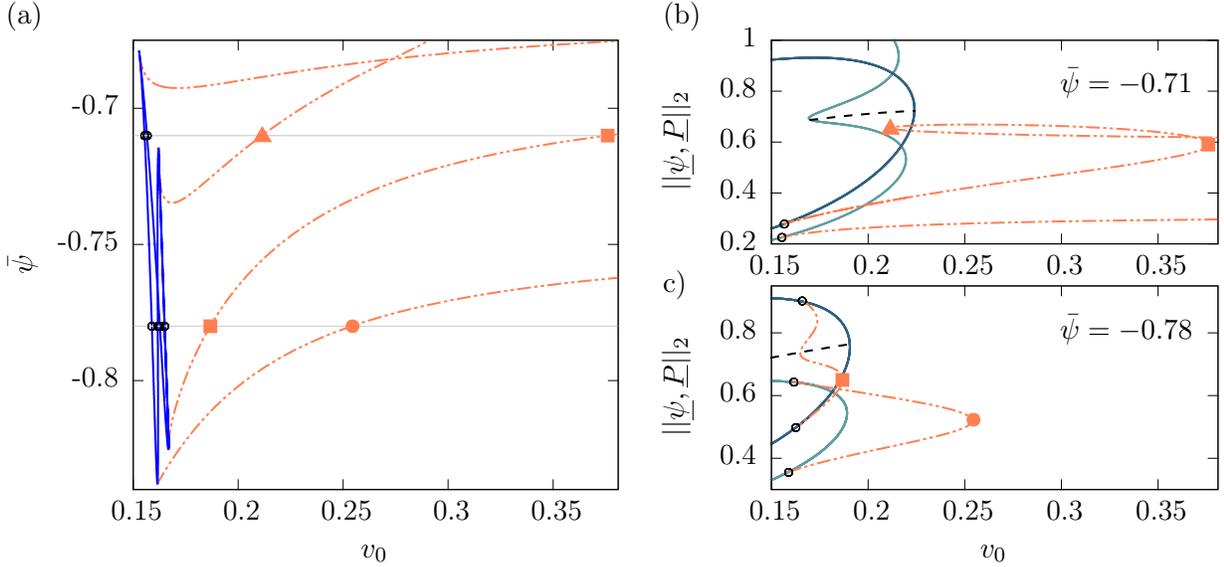}
\caption{\label{fig:fold-cont}(a) Two parameter continuation of the loci of the drift bifurcations (blue solid lines) and of the saddle-node bifurcations (orange  dot-dashed lines) of the one- and two-peak TLS. (b) and (c) Corresponding one parameter bifurcation diagrams at fixed values of $\bar{\psi}$ marked by gray horizontal lines in (a). Blue branches correspond to RLS, and dot-dashed orange branches correspond to TLS. The drift bifurcations are marked by circles, and the saddle-node bifurcations are indicated by orange symbols. For increasing mean concentrations the interval of $v_0$ in which moving LS exist (onset of motion up to fold) grows and ultimately the activity value at the fold diverges, i.e., TLS exist for arbitrarily high activities.}
\end{figure*}
 
A two-parameter continuation allows one to track the loci of various bifurcation points in a two parameter plane \cite{DoKK1991ijbc}. Here, we follow the loci of (i) the saddle-node bifurcations that mark the points where stable and unstable  one-peak and two-peak TLS annihilate and (ii) the drift bifurcations where TLS emerge from RLS in the parameter plane spanned by activity $v_0$ and mean density $\bar\psi$. This allows us to determine the area of existence of these localized states in the $(v_{0},\bar{\psi})$-plane.

The result is displayed in Fig.~\ref{fig:fold-cont}(a) where drift and saddle-node bifurcations are marked by blue solid lines and orange dot-dashed lines, respectively. The plot has to be carefully interpreted as the various bifurcations can be located on different branches in the bifurcation diagrams. To facilitate this we have marked the two values of $\bar{\psi}=-0.71$ and $\bar{\psi}=-0.78$ by horizontal gray lines and provide the corresponding one parameter bifurcation diagrams as Fig.~\ref{fig:fold-cont} (b) and (c) (also cf.~Fig.~\ref{fig:bifv0}), where the bifurcation points are highlighted by symbols, that also mark the fold continuation lines in (a).

Fig.~\ref{fig:fold-cont} proves that traveling localized states are a generic solution of the active PFC model as they occur in an extended region of the parameter plane. 
In fact, the values of $v_0$ at the saddle-node bifurcations that limit their existence diverge at $\bar{\psi}=-0.74$ and $\bar{\psi}=-0.69$ for one- and two-peak TLS, respectively. We numerically follow their position up to $v_0\gtrsim10^3$. Note that for $\bar{\psi}=-0.71$ the fold of the one-peak TLS has already moved far outside of the displayed $v_0$-interval. At this $\bar{\psi}$, the two-peak TLS exists up to $v_0\approx0.38$ while at $\bar{\psi}=-0.78$ its range of existence is smaller. All drift bifurcations are quite close to $v_0=0.15$ with only small variations between different localized states and with $\bar{\psi}$. This makes an interpretation of the corresponding diagram region challenging.

Roughly speaking, one-peak [two-peak] TLS exist in the lower part of Fig.~\ref{fig:fold-cont}(a) in the area between the nearly vertical blue solid lines and the dot-dashed line marked by the filled circle [square] while in the upper part of Fig.~\ref{fig:fold-cont}(a) they exist in the area between the dot-dashed line marked by the filled triangle and the one marked by the filled circle [square]. Remember that in (b) the filled circle has left the displayed interval of $v_0$. The uppermost unmarked dot-dashed line in Fig.~\ref{fig:fold-cont}(a) is related to three-peak TLS and will be further discussed elsewhere. 

\subsection{\label{subsec:linstab}Linear stability}

\begin{figure}
\includegraphics{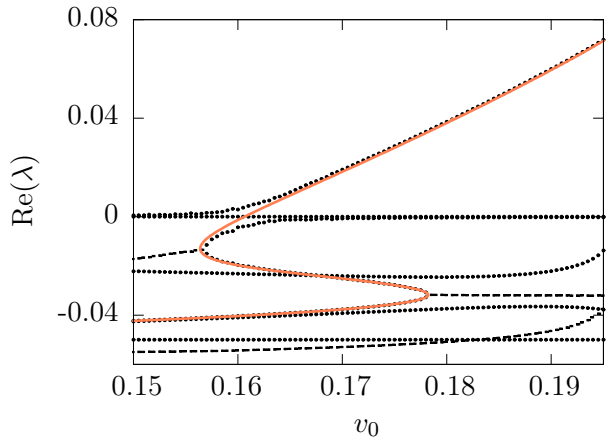}
\caption{\label{fig:EW} Black lines: Real part of eigenvalues obtained from numerical LSA with a finite difference method. The black dashed [dotted] lines indicate a complex [real] eigenvalues. Orange line: Real eigenvalue from continuation. As expected, two neutrally stable modes with Re($\lambda$) = 0 are found (translation mode and volume mode). One mode is destabilized at $v_c \approx 0.161$, the detected onset of motion. At $v_c$ the mode coincides with the spatial derivative of the localized state and corresponds to a translation.}
\end{figure}

\begin{figure*}
\includegraphics{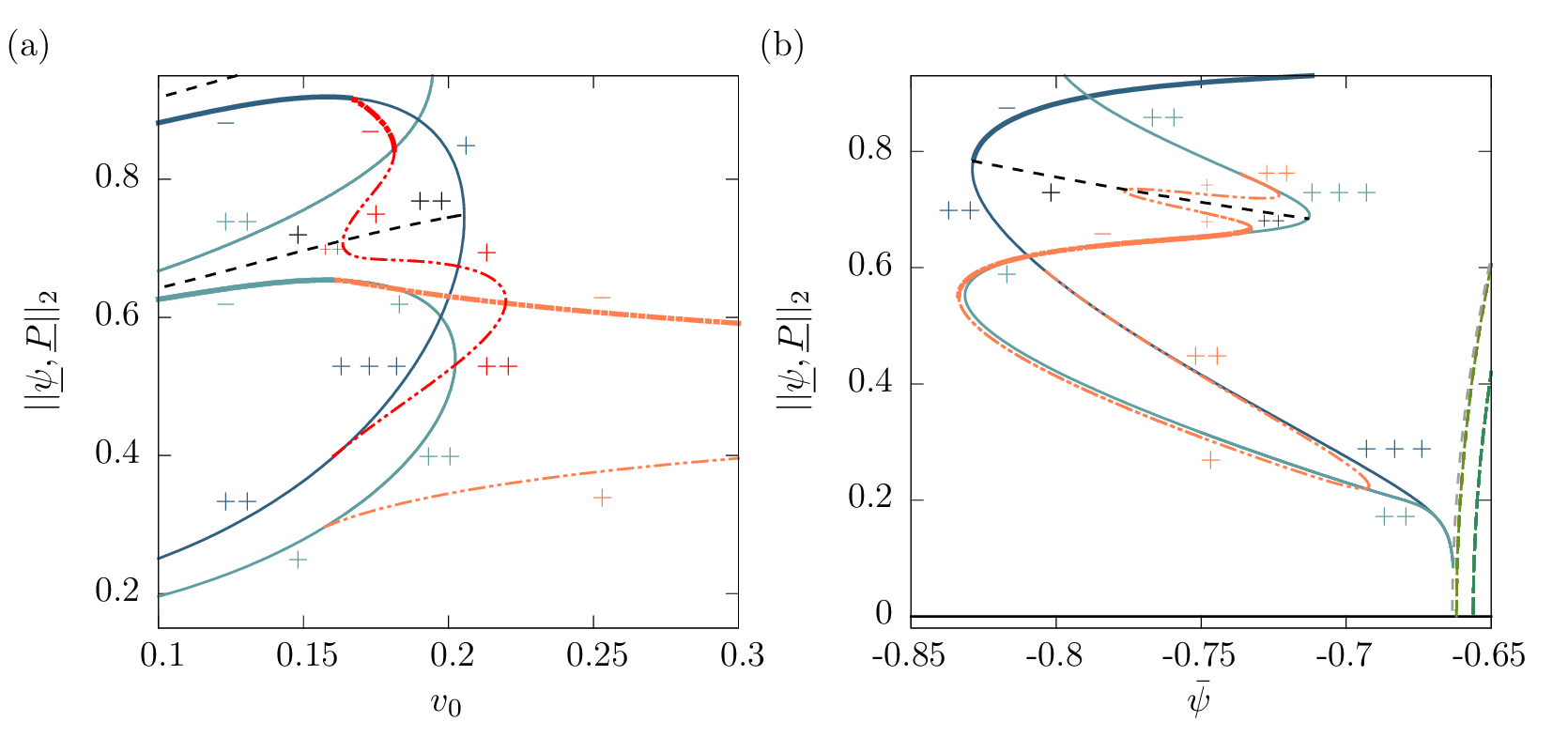}
\caption{\label{fig:stability}Linear stability of localized states. Light and dark blue lines: $\mathrm{RLS_{\mathrm{odd}}}$ and $\mathrm{RLS_{\mathrm{even}}}$. Dashed black: Asymmetric RLS. Dot-dashed orange: TRS. Stable steady states are indicated by $-$ signs and plotted as heavy lines. For unstable states, the number of $+$ signs gives the number of unstable eigenmodes with $\mathrm{Re}(\lambda)>0$. (a) Continuation of $v_0$, $\bar{\psi}=-0.75$. Symmetric RLS lose their stability in drift bifurcations at $v_0\approx0.16$ and TLS become stable. Asymmetric RLS are always unstable. (b) Continuation of $\bar{\psi}$, $v_0=0.16475$. LS are created in a subcritical bifurcation, branching off from the periodic branch (dashed gray, more periodic branches in dashed green). Note that vertical cuts at the respective values of $v_0$ of (b) in (a) and of $\bar{\psi}$ vice versa correspond to each other.}
\end{figure*}

Up to here we have discussed bifurcation diagrams and existence of solutions. Although general knowledge about bifurcations allows one to develop quite a good idea about the stability of the various solutions, it is important to explicitly determine the linear stability. The obtained detailed information then permits us to predict which states can persist in experiments or direct numerical simulations (the linearly stable states) and which states  may only appear as (possible long-lived) transients. These are given by the unstable states that represent saddles in function space, as they might first attract time-evolutions to then repel them into well defined directions corresponding to the eigenvectors of the most unstable eigenvalue.

For the analysis, Eqs.~(\ref{eq:dtpsi}) and (\ref{eq:dtP}) are linearized in small perturbations  $\delta\psi$ and $\delta P$ about a one-dimensional steady state $\left(\psi_{0}(x),P_{0}(x)\right)^T$ to obtain
\begin{align}
\partial_{t}\delta\psi &= \partial_{xx}\left(\epsilon+3(\bar{\psi}+\psi_{\mathrm{0}})^{2}+\left(1+\partial_{xx}\right)^{2}\right)\delta\psi-v_{0}\partial_{x}\delta P, \label{eq: lin dt psi} \\
\partial_{t}\delta P &= \partial_{xx}\left(C_{1}\delta P\right)-D_{\mathrm{r}}C_{1}\delta P-v_{0}\partial_{x}\delta\psi. \label{eq: lin dt P}
\end{align}
In the case of uniformly moving states $\left(\psi_{0},P_{0}\right)^T=\left(\psi_{0}(x+ct),P_{0}(x+ct)\right)^T$, a comoving frame term is added to the right-hand side.
Assuming exponential growth of the perturbation, i.e., $\delta\psi=\psi_1\exp(\lambda t)$ and $\delta P=P_1\exp(\lambda t)$ we have to solve
the linear eigenvalue problem: 
\begin{align}
\mathcal{L}\left(\psi_{0},P_{0}\right)\left(\begin{array}{c}
\psi_1\\
P_1
\end{array}\right)=\lambda\left(\begin{array}{c}
\psi_1\\
P_1 \label{eq:linprob}
\end{array}\right),
\end{align}
where the linear operator $\mathcal{L}\left(\psi_{0},P_{0}\right)$ is defined by the right-hand side of Eqs.~(\ref{eq: lin dt psi}) and (\ref{eq: lin dt P}) [it is explicitly given below in Eq.~(\ref{eq: linoperator})]. 

We are not able to pursue an analytical solution of the linear problem because already the steady states $\psi_{0}(x)$ and $P_{0}(x)$ are obtained by numerical continuation. Instead, we discretize the steady states equidistantly in space, i.e., employ a finite difference method to transform (\ref{eq:linprob}) into a standard linear algebraic eigenvalue problem that we solve employing standard numerical routines. 

The black lines in Fig.~\ref{fig:EW} give an example of a calculated eigenvalue spectrum in dependence of the activity. Shown are the real parts of the leading ten eigenvalues for the branch of one-peak RLS that in Fig.~\ref{fig:stability} is stable at $v_0 =0.1$. The dotted lines indicate purely real eigenvalues whereas the three dashed lines indicate complex eigenvalues. The largest eigenvalue is real and crosses zero at a critical activity of $v_c\approx0.161$ where the drift-pitchfork bifurcation occurs, as discussed in detail in section~\ref{sec:level1drift}. The obtained $v_c$ well agrees with the value we obtain through the numerical continuation of the one-peak TLS that detects the drift-pitchfork bifurcation (as a fold) at the same value. Note that in the discretized eigenvalue problem the zero crossing has to be obtained by extrapolation as the relevant eigenvalue 'interacts' with one of the two zero eigenvalues, in this way 'blurring' the crossing. This is related to the problem of level repulsion or avoided crossing (von Neumann-Wigner theorem~\cite{NWT1929}). To prevent the blurred zero crossing, we also solve Eq.~(\ref{eq:linprob}) by numerical continuation \cite{cenosTutorialLindrop}. The eigenvalue we obtain in this way is given by the orange line in Fig.~\ref{fig:EW}. It confirms the finite difference calculations and perfectly matches $v_c$.

Two zero eigenvalues exist for all $v_0$ and represent neutrally stable modes that are related to the symmetries of the model. One of them represents a translation mode that occurs because Eqs.~(\ref{eq:dtpsi}) and (\ref{eq:dtP}) are invariant with respect to translation and, therefore, exhibit the neutral eigenmode of translation, often called Goldstone mode of translational symmetry. In addition, an infinitesimal change in the mean concentration $\bar{\psi}$ does also result in another solution of the equations, i.e., the second mode with zero eigenvalue is a neutral volume mode or Goldstone mode of symmetry with respect to mass change.

Calculating the eigenfunction that is destabilized shows that at $v_c$ the mode matches the spatial derivative of the investigated localized peak. The derivative corresponds to an infinitesimal shift of the position of the peak and, therefore, to the Goldstone mode of translational symmetry. This fact indicates that the onset of motion is indeed due to a drift bifurcation. 

A typical result of a systematic stability analysis is shown in Fig.~\ref{fig:stability}, where (a) represents an enlargement off a part of the bifurcation diagram in Fig.~\ref{fig:bifv0} and (b) is the lowest part of the snakes-and-ladders structure. The branches of linearly stable and unstable states are indicated by ``-'' and ``+'' signs, respectively. The number of ``+'' signs gives the number of unstable eigenmodes.
Linearly stable states are represented by heavy lines, indicating that in (a) in the considered parameter range one has stable one- and two-peak RLS and TLS with regions of multistability of (i) one- and two-peak RLS at low activity, (ii) one- and two-peak TLS at slightly larger activity and in between (iii) a very small region where one-peak TLS and two-peak RLS are both linearly stable. In the considered case all the eigenvalues that cross the imaginary axis are real, although stable complex eigenvalues do occur (see dashed lines in Fig.~\ref{fig:EW}). Note that Fig.~\ref{fig:stability}(a) shows more bifurcations than are followed in Fig.~\ref{fig:fold-cont}(a).

Studying Figs.~\ref{fig:stability}(b) and \ref{fig:snake-ladder} in detail one finds that - despite the similar shape of the snake and ladder - the stability of the RLS differs from the one found for the PFC model \cite{TARG2013pre}: there the symmetric RLS change their stability as the branches snake along, while the asymmetric RLS are always unstable. Here, however, the stable symmetric RLS are already destabilized before the saddle-node bifurcation is reached as the TLS emerge at the drift-pitchfork bifurcation, i.e., their range of linear stability is diminished. Since $v_0=0.16475>v_c$ in (b) most of the resting branches are unstable. At a drift-transcritical bifurcation the asymmetric RLS also acquire an additional unstable mode as compared to the case of PFC. For activities lower than $v_c$ the picture is very similar to the one of the passive PFC model. 

\begin{figure}
\includegraphics{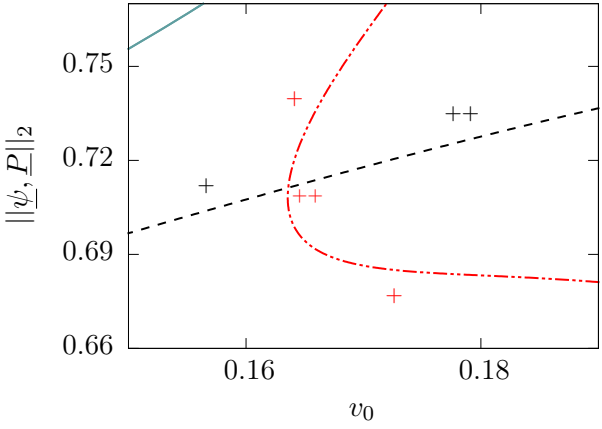}
\caption{\label{fig:transcritical_bif}Drift-transcritical bifurcation. Enlargement of Fig.~\ref{fig:stability}(a). The asymmetric RLS (dashed black, +) acquires an additional unstable mode (++) in a drift-transcritical bifurcation. The moving double bump (dot-dashed red line) changes its stability in the transcritical bifurcation and at the nearby fold. All shown branches are linearly unstable.}
\end{figure}

Figure \ref{fig:transcritical_bif} enlarges a detail of Fig.~\ref{fig:stability}(a): the drift-transcritical bifurcation, where moving states branch off the asymmetric resting state composed of two density peaks of different height. As already the resting state is asymmetric, the two sub-branches emerging at the drift bifurcation are not related  to each other by symmetry, but intrinsically differ. Hence, in this case the creation of the TLS corresponds to a drift-transcritical bifurcation, different from the drift-pitchfork bifurcations in which the symmetric RLS lose their stability. The transcritical bifurcation does not coincide with the fold of the (red) TLS branch and its stability changes twice close to the drift bifurcation. Accordingly, in Fig.~\ref{fig:stability}(a) the two sub-branches of TLS seem to have the same stability before and after crossing the resting asymmetric state. There is another the drift-transcritical bifurcation on the asymmetric branch in Fig.~\ref{fig:stability}(b).

\section{\label{sec:level1drift}Onset of Motion - the Drift Instability}

Next we discuss the numerically found drift bifurcations more in detail and derive a specific simple analytic condition that allows one to detect drift bifurcations for a class of models that includes the aPFC model. The analytical criterion for the onset of motion is valid for the encountered drift-pitchfork and drift-transcritical bifurcations.

\subsection{Velocity expansion}
We consider the one-dimensional version of the model (\ref{eq:dtpsi}) and (\ref{eq:dtP}) in a comoving frame with coordinate $x'=x+ct$, time $t$ and velocity $c$. We use  $\left(\psi_{0}(x),P_{0}(x)\right)^T$ to denote a steady solution, i.e., with $c=0$. Assuming there are only small corrections
$(\tilde{\psi}_i,\tilde{P}_i)^T$ to the steady state
when changing parameters close to the drift bifurcation, we introduce a velocity expansion 
\begin{align}
\psi &= \psi_{0}(x)+c\left[\tilde{\psi}_{1}(x)+c\tilde{\psi}_{2}(x)+c^{2}\tilde{\psi}_{3}(x)+\ldots\right],\\
P &= P_{0}(x)+c\left[\tilde{P}_{1}(x)+c\tilde{P}_{2}(x)+c^{2}\tilde{P}_{3}(x)+\ldots\right].\nonumber
\end{align}
Plugging in the expansions (up to order $c^{2}$) in the dynamic equations (Eqs.~(\ref{eq:dtpsi}) and (\ref{eq:dtP})) 
leads to
\begin{widetext}
\begin{align}
  c\,\partial_{x}\left(\psi_{0}+c\tilde{\psi}_{1}+c^{2}\tilde{\psi}_{2}\right) =& \partial_{xx}\left[\left(\epsilon+\left(1+\partial_{xx}\right)^{2}\right)\left(\psi_{0}+c\tilde{\psi}_{1}+c^{2}\tilde{\psi}_{2}\right)+\left(\bar{\psi}+\psi_{0}+c\tilde{\psi}_{1}+c^{2}\tilde{\psi}_{2}\right)^{3}\right],\nonumber \\
 &  -v_{0}\partial_{x}\left(P_{0}+c\tilde{P}_{1}+c^{2}\tilde{P}_{2}\right)\\
c\,\partial_{x}\left(P_{0}+c\tilde{P}_{1}+c^{2}\tilde{P}_{2}\right) =& \left(\partial_{xx}-D_{\mathrm{r}}\right)\left[C_{1}\left(P_{0}+c\tilde{P}_{1}+c^{2}\tilde{P}_{2}\right)+C_{2}\left(P_{0}+c\tilde{P}_{1}+c^{2}\tilde{P}_{2}\right)^{3}\right]\nonumber \\
 &  -v_{0}\partial_{x}\left(\psi_{0}+c\tilde{\psi}_{1}+c^{2}\tilde{\psi}_{2}\right).\nonumber
\end{align}
\end{widetext}
\newpage{}
By equating coefficients of $c^{n}$, we find for $c^{0}$ 
\begin{align}
0 &= \partial_{xx}\left[\left(\epsilon+\left(1+\partial_{xx}\right)^{2}\right)\psi_{0}+\left(\bar{\psi}+\psi_{0}\right)^{3}\right] -v_{0}\partial_{x}P_{0}\\
0 &= \left(\partial_{xx}-D_{\mathrm{r}}\right)\left(C_{1}P_{0}+C_{2}P_{0}^{3}\right)-v_{0}\partial_{x}\psi_{0},\nonumber
\end{align}
i.e., we recover the equations for the resting base state. To linear order in $c$ we obtain
\begin{align}
\partial_{x}\psi_{0}  =&    \partial_{xx}\left[\left(\epsilon+\left(1+\partial_{xx}\right)^{2}\right)\tilde{\psi}_{1}+3\left(\bar{\psi}+\psi_{0}\right)^{2}\tilde{\psi}_{1}\right]\nonumber\\
& -v_{0}\partial_{x}\tilde{P}_{1} \label{eq:linear in c-1}\\
\partial_{x}P_{0} =& \left(\partial_{xx}-D_{\mathrm{r}}\right)\left(C_{1}\tilde{P}_{1}+3\,C_{2}P_{0}^{2}\tilde{P}_{1}\right)-v_{0}\partial_{x}\tilde{\psi}_{1}.
\nonumber 
\end{align}
i.e., a linear system for $\tilde{\psi}_{1}$ and $\tilde{P}_{1}$. We write Eqs.~(\ref{eq:linear in c-1}) in matrix form
\begin{align}
\partial_{x}\left(\begin{array}{c}
\psi_{0}\\
P_{0}
\end{array}\right) &= \mathcal{L}(\psi_{0},P_{0})\left(\begin{array}{c}
\tilde{\psi}_{1}\\
\tilde{P}_{1}
\end{array}\right)\label{eq:phi0=00003DL'phi1-1}
\end{align}
with the same linear operator $\mathcal{L}$ already employed in (\ref{eq:linprob}):
\begin{widetext}
\begin{equation}
\mathcal{L}(\psi_{0}(x),P_{0}(x))=\left(\begin{array}{cc}
\partial_{xx}\left[\left(\epsilon+\left(1+\partial_{xx}\right)^{2}\right)+3\left(\bar{\psi}+\psi_{0}(x)\right)^{2}\right] & -v_{0}\partial_{x}\\
-v_{0}\partial_{x} & \left(\partial_{xx}-D_{\mathrm{r}}\right)\left(C_{1}+3\,C_{2}P_{0}(x)^{2}\right)
\end{array}\right)
\label{eq: linoperator}.
\end{equation}
\end{widetext}

In the following, we focus again on the case of a linear equation for $P$ without spontaneous polarization, $C_2=0$. We notice that the top left component of (\ref{eq: linoperator})
\begin{align}
L_{11}(x) &= \partial_{xx}\left[\left(\epsilon+\left(1+\partial_{xx}\right)^{2}\right)+3\left(\bar{\psi}+\psi_{0}(x)\right)^{2}\right]\nonumber \\
 &= \partial_{xx}\, L_{\mathrm{SH}}(\psi_{0}(x))
\end{align}
is the product of a Laplacian (due to mass conservation) and the linearized operator from a Swift-Hohenberg equation with cubic nonlinearity. This fact will turn out to be very helpful when forming the adjoint operator $\mathcal{L}^{\dagger}$.
\subsection{Translational symmetry and Goldstone modes}
\lo{Adding its first spatial derivative to the base state corresponds to a small shift in the position of the state. Since the aPFC model is translationally invariant, 
\begin{align}
\partial_{x}\left(\begin{array}{c}
\psi_{0}\\
P_{0}
\end{array}\right) &= \left(\begin{array}{c}
\mathcal{G}_{1}\\
\mathcal{G}_{2}
\end{array}\right)\equiv\left(\begin{array}{c}
\psi_{\mathcal{G}}\\
P_{\mathcal{G}}
\end{array}\right)
\end{align}
can be identified as a neutral eigenfunction with eigenvalue zero, often referred to as the Goldstone mode $\mathcal{G}$ of the translational symmetry. Thus,
\begin{equation}
\mathcal{L}\,\partial_{x}\left(\begin{array}{c}
\psi_{0}\\
P_{0}
\end{array}\right)=\mathcal{L} \, \mathcal{G}=\mathbf{0}.
\label{eq:goldtrans}
\end{equation}}
A typical destabilization occurs when the real part of an eigenvalue crosses zero as parameters of the system are being changed. We now consider the case that the imaginary part also equals zero, so that the corresponding eigenfunctions of $\mathcal{L}$ can be expressed by a linear combination of the Goldstone modes. The second Goldstone mode mentioned in Section~\ref{subsec:linstab} is the volume mode that does not interfere in the drift bifurcation. At the bifurcation point, a real eigenvalue crosses the imaginary axis, i.e., an additional neutral mode exists. In consequence, the system of eigenfunctions of the null space of the linear operator is incomplete and must be supplemented by a generalized neutral eigenfunction \cite{driftbif_gurevich}. This function is called the propagator mode $\mathcal{P}$, defined by 
\begin{equation}
\mathcal{L}\mathcal{P}=\mathcal{G}.\label{eq:LP=00003DG propagator mode-1}
\end{equation}
It is exactly the occurrence of $\mathcal{P}$ that marks the destabilization, i.e., the onset of motion. Using the Fredholm alternative \cite{evans}, one finds that Eq.~(\ref{eq:LP=00003DG propagator mode-1}) can be solved iff 
\begin{equation}
\langle\mathcal{G}^{\dagger}|\mathcal{G}\rangle=0,\label{eq:<G+,G>=00003D0}
\end{equation}
where $\mathcal{G}^{\dagger}$ is the neutral eigenfunction of the adjoint operator $\mathcal{L}^{\dagger}$ with the same spatial symmetry as $\mathcal{G}$. The scalar product $\langle\cdot|\cdot\rangle$ is defined as a full spatial integration over the considered domain. The values of a set of control parameters for which Eq.~(\ref{eq:<G+,G>=00003D0}) is fulfilled corresponds to the bifurcation point.
\subsection{The adjoint linearized operator}
Let $\mathcal{G}^{\dagger}$ be the adjoint neutral eigenfunction,
i.e., 
\begin{equation}
\mathcal{L}^{\dagger}\,\mathcal{G}^{\dagger}=\mathbf{0}.
\label{eq:anef}
\end{equation}
Equation~(\ref{eq:phi0=00003DL'phi1-1}) corresponds to
\begin{equation}
\mathcal{G} = \mathcal{L}\left(\begin{array}{c}
\tilde{\psi_{1}}\\
\tilde{P_{1}}
\end{array}\right)
\end{equation}
showing that $(\tilde{\psi}_{1},\tilde{P}_{1})^T$ is a generalized neutral eigenfunction $\mathcal{P}$. To find $\mathcal{G}^{\dagger}=(\psi_{\mathcal{G}}^{\dagger},P_{\mathcal{G}}^{\dagger})^T$ we determine the adjoint operator
\begin{align}
\mathcal{L}^{\dagger} &= \left(\begin{array}{cc}
L_{\mathrm{SH}}\,\partial_{xx} & v_{0}\partial_{x}\\
v_{0}\partial_{x} & C_{1}\left(\partial_{xx}-D_{\mathrm{r}}\right)
\end{array}\right)
\end{align}
using $(AB)^{\dagger}=B^{\dagger}A^{\dagger}$, the self-adjointness of $\partial_{xx}$
and $L_{\mathrm{SH}}$, the relation $\partial_{x}^{\dagger}=-\partial_{x}$,
and ($v_{0}, C_{1}, D_{\mathrm{r}})\in\mathbb{R}$.
\subsection{Determining the adjoint eigenfunctions}
\lo{
Componentwise the adjoint problem reads 
\begin{align}
0 &= L_{\mathrm{SH}}\,\partial_{xx}\psi_{\mathcal{G}}^{\dagger}+ v_0\partial_xP_{\mathcal{G}}^{\dagger} \label{eq:adjoint1}\\
0 &= v_0\partial_x\psi_{\mathcal{G}}^{\dagger} + C_1\left(\partial_{xx}-D_{\mathrm{r}}\right)P_{\mathcal{G}}^{\dagger}\label{eq:adjoint2}
\end{align}

Comparing Eq.~(\ref{eq:adjoint1}) to the steady state equation for $\psi$ (\ref{eq:steadystatePSI}) with $J=0$ and $c=0$ and employing a simple chain rule
\begin{align}
 0 &= \partial_x \frac{\delta\mathcal{F}}{\delta\psi}(\psi_0)-v_0P \\
  &= L_{\mathrm{SH}}\partial_x\psi_0-v_0P
\end{align}
suggests 
\begin{align}
\partial_{xx}\psi_{\mathcal{G}}^{\dagger} &= \partial_{x}\psi_{0},\\
\partial_{x}P_{\mathcal{G}}^{\dagger} &= -P_{0}
\end{align}
Integrating yields
\begin{align}
\psi_{\mathcal{G}}^{\dagger}(x)&=\int_{0}^{x}\left(\psi_{0}(x')+\mathrm{C}\right)\mathrm{d}x'+\mathrm{D},\\
P_{\mathcal{G}}^{\dagger}(x)&=-\int_{0}^{x}P_{0}(x')\mathrm{d}x'+\mathrm{F}
\label{eq:integratedtwice}
\end{align}
with constants C, D, F. Eq.~(\ref{eq:adjoint2}) is consistent with this neutral adjoint eigenfunction. Substituting gives 
\begin{align}
 v_0\psi_0-C_1(\partial_{xx}-D_\mathrm{r})\int P_{0}(x)\mathrm{d}x=\mathrm{const.}
\end{align}
which is true as can be seen by integrating the steady state equation for $P$, Eq.~(\ref{eq:steadystateP}).
}

\begin{figure}
\includegraphics{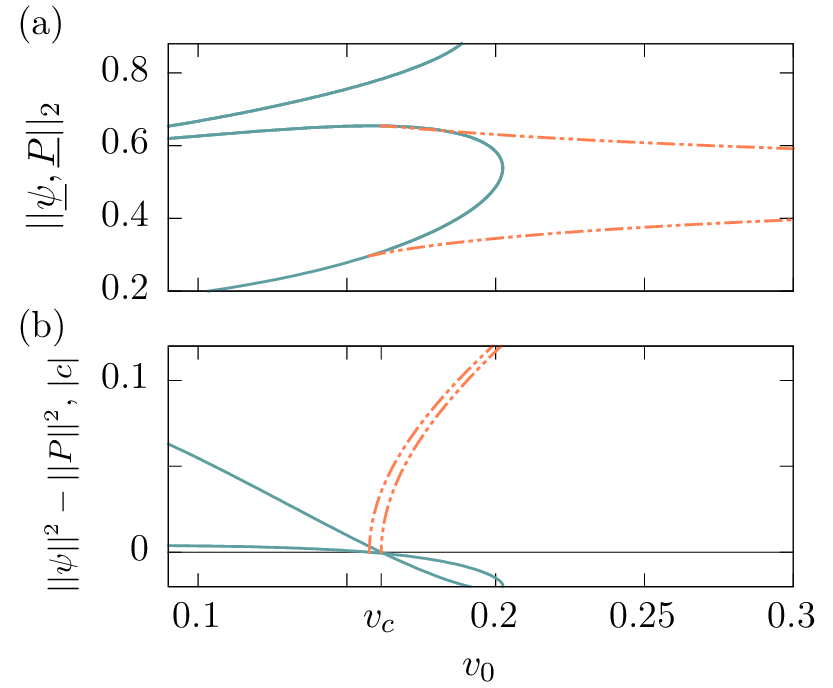}
\caption{\label{fig:fredholm}Onset of motion. (a) $L^2$-Norm of steady states in dependence of $v_0$ for fixed $\bar{\psi}=-0.75$. The blue branch corresponds to a RLS with one bump. The RLS is destabilized at $v_c$ and starts to travel with drift velocity $c$. The traveling odd LS is indicated by the dashed orange branch. (b) Solvability condition Eq.~(\ref{eq:normdiff}) $||\psi_{0}||^2-||P_{0}||^2$ (blue) of the RLS and velocity $|c|$ (dashed orange line) of TLS vs. activity $v_0$ showing perfect agreement of the two approaches.}
\end{figure}

\subsection{Solvability condition}
Collecting all the results, the solvability condition (\ref{eq:<G+,G>=00003D0}) reads 
\begin{align}
  \langle\mathcal{G}^{\dagger}|\mathcal{G}\rangle &= \langle\psi_{\mathcal{G}}^{\dagger}|\psi_{\mathcal{G}}\rangle+\langle P_{\mathcal{G}}^{\dagger}|P_{\mathcal{G}}\rangle\nonumber \\
  &= \langle\int_{0}^{x}\left(\psi_{0}(x')+\mathrm{C}\right)\mathrm{d}x'+\mathrm{D}|\partial_{x}\psi_{0}\rangle \nonumber\\
  &\;\;-\; \langle\int_{0}^{x}\left(P_{0}(x')+\mathrm{E}\right)\mathrm{d}x'|\partial_{x}P_{0}\rangle \\
  &= -\langle\partial_{x}\left(\int_{0}^{x}\left(\psi_{0}(x')+\mathrm{C}\right)\mathrm{d}x'+\mathrm{D}\right)|\psi_{0}\rangle\nonumber \\
  &\;\;+\; \langle\partial_{x}\int_{0}^{x}\left(P_{0}(x')+\mathrm{E}\right)\mathrm{d}x'|P_{0}\rangle  \\
  &= -\langle\psi_{0}|\psi_{0}\rangle+\langle P_{0}|P_{0}\rangle=0. \\
  \Leftrightarrow0 &= ||\psi_{0}||^{2}-||P_{0}||^{2}\label{eq:normdiff},
\end{align}
where we have employed a partial integration and used $\langle\mathrm{C}|\psi_{0}\rangle=\mathrm{C}\int_{0}^{L}\psi_{0}\mathrm{d}x=0$, since $\psi$ is the modulation around the fixed mean density. The same holds for the integral over $P_0$ as explained in Section~\ref{sec:level2}. For all TLS we have found the onset of motion perfectly matches the zero crossing of $||\psi_{0}||^2-||P_{0}||^2$.

A particular example is given in Fig.~\ref{fig:fredholm}. Panel (a) shows a part of the bifurcation diagram Fig.~\ref{fig:bifv0}. The solid blue branch corresponds to the RLS with a single density peak that loses its stability in a drift-pitchfork bifurcation at $v_c$. The emerging traveling bump (upper dot-dashed orange line) is linearly stable (cf. Fig.~\ref{fig:stability}(a), the lower orange branch is unstable). In the lower panel, Fig.~\ref{fig:fredholm}(b), we plot the difference of the squared norms as employed in Eq.~(\ref{eq:normdiff}). In addition, we also display the velocity of the emerging TLS (dot-dashed orange). The two zero crossings of $||\psi_{0}||^2-||P_{0}||^2$ occur at exactly the same values of $v_0$ as the onsets of motion. The second root is due to the lower unstable branch of TLS that bifurcates at a slightly lower activity. Notice that the criterion for the onset of motion, Eq.~\ref{eq:normdiff}, also holds for the drift-transcritical bifurcation.

\section{\label{sec:level1con}Discussion and conclusions}

We have in some detail studied the bifurcation structure of the active Phase-Field-Crystal model in the one-dimensional case. \lo{After discussing the linear stability of the liquid (homogeneous) state with respect to monotonous and oscillatory modes, we have briefly discussed the existence and stability of stable domain-filling resting and traveling crystalline (periodic) structures. Note that we have not systematically studied unstable domain-filling periodic structures. Our main focus has been on crystallites (crystals of finite extension) that correspond to stable and unstable localized states of different symmetries. We have analyzed how the classical slanted snakes-and-ladders structure (slanted homoclinic snaking) known from the Phase-Field-Crystal model \cite{TARG2013pre} is amended by activity. In particular,} we have shown that increasing activity, one finds a critical value for the onset of motion of the various localized states and of the domain-filling crystal. Using the mean concentration $\bar\psi$ as control parameter we have found that traveling states emerge either through drift-pitchfork bifurcations of the resting parity (left-right) symmetric localized states or through drift-transcritical bifurcations of resting asymmetric localized states that form the rungs of the snakes-and-ladders bifurcation structure. At the studied parameter values these traveling localized states always occur within the $\bar\psi-$range limited by the snaking branches of resting localized states.

Note that this onset behavior differs from the case of the non-variational Swift-Hohenberg equations studied in Refs.~\cite{HoKn2011pre}. There, at any value of the driving parameter in front of the non-variational term all asymmetric states drift and the original pitchfork bifurcations of the variational system either split into two saddle-node bifurcations or become a drift-pitchfork bifurcation. Here, however, the coupling of the two fields allows for resting asymmetric states even for finite activity parameter and moving states emerge through drift bifurcations that are not present (in any form) in the variational limit.

The second investigated main control parameter has been the activity. Here, the general tendency is that an increase in activity suppresses the resting localized and periodic states that ultimately annihilate in saddle-node bifurcations at critical activities that are of a similar magnitude for all studied states. \lo{In other words activity ultimately melts all resting crystalline structures as the driving force overcomes the attractive forces that stabilize the equilibrium crystals and crystallites that exist in the reference system without activity. This corresponds to the melting of equilibrium clusters by activity observed in the Brownian dynamics simulations of Ref.~\cite{ReBH2013pre} for self-propelled particles with short-range attraction.}
However, at values of the activity below this melting point most branches of resting states show drift bifurcations where branches of traveling states emerge that may exist in a small range of activity or even extend towards infinite activity as we have shown by numerical two-parameter continuation of the relevant bifurcations. In other words, depending on parameters, although activity may melt traveling crystallites, there are extended parameter regions where this is not the case. \lo{In fact, we have found that although a high activity melts most traveling localized states, i.e., traveling crystalline patches, this is not the case for traveling periodic states, i.e., traveling domain-filling crystals. They can be driven with arbitrarily high activity and then show high velocities. We believe, that this is most likely the case because the periodicity of the domain-filling crystals is fixed, while the traveling localized states naturally adapt their peak spacing. This additional degree of freedom could make them less stable. Note that the found crystallites are unrelated to the motility-induced clusters discussed, e.g., in \cite{Ginot2015prx,SSWK2015prl,CaTa2015arcmp}. The latter effect has not yet been found in an active PFC model as they are mainly considered to study how equilibrium crystallization is amended by activity. It should be further investigated whether it may also describe motility-induced clustering, especially when allowing for spontaneous polarization ($C_2\neq0$).}

Furthermore, we have investigated the region of existence of traveling localized states and have shown that they are generic solutions for extended regions of the plane spanned by mean concentration and activity. Whereas extended traveling localized states of three and more peaks quickly vanish into the homogeneous background, narrow localized states (one and two density peaks) can be driven at quite high activities where they reach high velocities. This does not seem to be the case in the \lo{nonvariational} systems studied in \cite{HoKn2011pre,BuDa2012sjads}. Therefore, a comparative study of the present system, the systems studied in \cite{HoKn2011pre,BuDa2012sjads} and the ones reviewed and discussed in \cite{KoTl2007c} would be beneficial.

A further focus has been the onset of motion that occurs at a critical activity which only slightly depends on the particular localized state. We have considered drift instabilities for the system of two coupled equations where one represents a mass-conserving dynamics of a density-like quantity and the second one is a linear equation for the polarization. Also the non-variational coupling of the two equations is linear. Under these conditions we have derived a general criterion for the onset of motion. Namely, the zero crossing of the difference
of the squared norms of the two steady fields ($||\psi_{0}||^2-||P_{0}||^2$) marks the onset of motion for all localized and extended crystalline states. The criterion holds for both types of drift instabilities that occur in the aPFC model: drift-pitchfork and drift-transcritical bifurcations and may be used to determine the critical strength of activity that is needed for collective traveling states. Note, that the criterion also applies to other models of active media that fulfill the described conditions. This will be discussed elsewhere. What needs further clarification is the question of whether such a simple criterion can be derived for more complicated active models, that do more faithfully model specific properties of the experimental systems. 

Finally, we highlight a number of questions that merit further investigation. \lo{Here, our main aim has been to establish a first overview of the rather involved overall bifurcation structure
that is related to the onset of motion in continuum models of active crystals. Although we have focused on a one-dimensional systems we believe} that most of the obtained results will hold for two- or even three-dimensional systems. There, however, the picture is complicated by the possible occurrence of various pattern types, compare, for instance, the differences found in the classical non-conserved Swift-Hohenberg model \cite{BuKn2006pre,ALBK2010sjads,LSAC2008sjads}. 
Having established the existence of the various traveling and resting localized states it will be interesting to study their interactions (in analogy to section~IV of Ref.~\cite{HoKn2011pre}), and to obtain more detailed information about their regions of existence, multistability and instabilities. As experimental studies often focus on the collective behavior of many interacting clusters \cite{theurkauff2012prl,Ginot2015prx,ginot2018aggregation}, it should be investigated whether it is possible to derive statistical models from single cluster bifurcation studies as the present one. Such a methodology has recently been presented for ensembles of sliding drops \cite{WTEG2017prl}.
\lo{We hope that the provided study will serve as a reference for other such analyses of more detailed models for active crystals, e.g., here we have focused on a rather simple coupling of concentration and polarization and have also excluded spontaneous polarization. The obtained results regarding the onset of motion should also be compared to related results regarding the bifurcation structure of other models of active matter. This will allow one to develop a clearer general understanding of observed multistabilities of states, hysteresis effects and thresholds where qualitative changes occur.}

\begin{acknowledgments}
We acknowledge support through the doctoral school ``Active living fluids'' funded by the German French University  (Grant No. CDFA-01-14). 
LO wishes to thank the foundation ``Studienstiftung des deutschen Volkes'' for financial support, Johannes Kirchner for fruitful discussions and Fenna Stegemerten and Tobias Frohoff-H\"ulsmann for their detailed feedback on the manuscript.
 
\end{acknowledgments}

% \bibliography{aPFC.bib}% Produces the bibliography via BibTeX.
\end{document}